\pdfoutput=1
\documentclass[twocolumn,prd,preprintnumbers,amsmath,amssymb,superscriptaddress]{revtex4}

\usepackage{graphicx}

\usepackage{dcolumn}
\usepackage{bm}
\usepackage{color}

\begin{document}


\title{Discriminating the source of high-energy positrons with AMS-02}
\author{Miguel Pato}
\email{pato@iap.fr}
\affiliation{Institute for Theoretical Physics, Univ.~of Z\"urich, Winterthurerst.~190, 8057 Z\"urich CH}
\affiliation{Institut d'Astrophysique de Paris, UMR 7095-CNRS, Univ.~Pierre \& Marie Curie, 98bis Bd Arago 75014 Paris, France}
\affiliation{Dipartimento di Fisica, Universit\`a degli Studi di Padova, via Marzolo 8, I-35131, Padova, Italy}
\author{Massimiliano Lattanzi}
\affiliation{International Centre for Relativistic Astrophysics and Dipartimento di Fisica, Universit\`a di Roma ``La Sapienza'', Piazzale A.~Moro 2, 00185, Rome, Italy}
\author{Gianfranco Bertone}
\affiliation{Institute for Theoretical Physics, Univ.~of Z\"urich, Winterthurerst.~190, 8057 Z\"urich CH}
\affiliation{Institut d'Astrophysique de Paris, UMR 7095-CNRS, Univ.~Pierre \& Marie Curie, 98bis Bd Arago 75014 Paris, France}

\date{\today}

\begin{abstract}
We study the prospects for discriminating between the dark matter (DM) and pulsar origin of the PAMELA positron excess with the Alpha Magnetic Spectrometer AMS-02. We simulate the response of AMS-02 to positrons (and electrons) originating from DM annihilations, and determine the pulsar parameters (spin-down luminosity, distance and characteristic age) that produce a satisfactory fit to the mock AMS-02 data. It turns out that it is always possible to mimic a DM signal with pulsars. Although the fit in some cases requires values of spin-down luminosity and characteristic age different from those of known pulsars in the ATNF and Fermi-LAT catalogues, these catalogues are known to be incomplete, and therefore the pulsar interpretation can hardly be ruled out. We also show that if the positron excess is due to a single pulsar, it is always possible to find a DM candidate that provides a good fit to the mock AMS-02 data. The discrimination between the two scenarios will thus require a better knowledge of the underlying sources, or complementary data.
\end{abstract}

\maketitle


\section{Introduction}\label{secintro}

\par The nature of high-energy cosmic ray electrons and positrons remains an open problem in modern astrophysics and it is currently a matter of intense debate. Data accumulated over the years led to the description of the electronic component in local cosmic rays as a single power-law with spectral index around 3.4 at multi-GeV energies \cite{Casadei}, where positrons are much less abundant than electrons. A standard picture has therefore emerged to explain the local flux of cosmic ray electrons and positrons in the context of galactic cosmic ray propagation. In this framework, positrons result from the inelastic scattering of cosmic ray protons and nuclei against the interstellar gas in our Galaxy (mainly H and He nuclei), and then diffuse and lose energy before arriving at the Earth. Electrons instead are believed to be primaries, presumably accelerated in supernova remnants \cite{Longair2} and injected into the interstellar medium (ISM). In the standard picture, the propagation of this injection spectrum would lead to the bulk of the local cosmic ray electrons.

\par Several hints, however, have challenged over the years the secondary nature of high-energy cosmic positrons, in particular the results of CAPRICE \cite{caprice94} and HEAT \cite{heat9495,heat00} that indicate a rather large positron fraction at multi-GeV energies. In the last few years, a host of observations of high-energy electrons and positrons has become available, including results from ATIC \cite{ATIC}, PPB-BETS \cite{PPBBETS}, PAMELA \cite{PAMELA,PAMELA10}, Fermi-LAT \cite{fermi,fermilat10} and H.E.S.S. \cite{hess08,hess09}. PAMELA, for instance, has measured a steeply rising positron fraction above $\sim$10 GeV and up to $\sim$100 GeV. Fermi-LAT results, on the other hand, show an electron plus positron flux of spectral index $\sim$3.0 above $\sim$20 GeV and with a possible hardening at about 300 GeV, while H.E.S.S.~hints at a cutoff of a few TeV. This so-called electron/positron excess is at odds with the standard picture described above. In fact, cosmic ray spallation on the galactic disk fails to produce enough positrons and, more importantly, is incompatible with a positron fraction that rises with increasing energies \cite{Delahaye08,Serpico}. As pointed out in \cite{Atoyan}, one needs one or more nearby and recent sources to accommodate the data. This is because $\sim$100 GeV positrons (and electrons) lose energy efficiently through inverse Compton scattering and synchrotron emission, presenting a cooling time of about $\sim 2\times 10^{6}$ yr which translates into kpc-scale diffusion distances. Several possible sources were put forward many years ago and recently revisited in the light of new electron-positron data.

\par One of the most popular hypotheses is to invoke dark matter (DM) annihilations \cite{cirelli} or decays \cite{Tran} in the galactic halo. Such annihilations or decays would produce high-energy electrons and positrons either directly or through the decay of secondary particles. Although exciting, this interpretation requires non-standard dark matter properties  -- such as high annihilation cross-sections \cite{Hisano:2004ds,Cirelli:2007xd,ArkaniHamed:2008qn,Pospelov:2008jd,Lattanzi:2008qa,MarchRussell:2008tu} -- that are in tension with other data, including radio emission and $\gamma$-rays \cite{Taoso,Profumo:2009uf,Belikov:2009cx}, or cosmic microwave background measurements \cite{galli,fink,Iocco}. Moreover, the DM interpretation requires annihilation mainly to leptonic channels in order to be consistent with the data on the antiproton flux, that itself is completely consistent with what is expected from secondary production \cite{cirelli,Delahaye2}. Another possible origin for the excess is the emission of electrons and positrons from mature pulsars \cite{Atoyan,Buesching,Hooper,Profumo,Delahaye3}. Indeed, electrons can be accelerated in the magnetosphere of pulsars and, due to the existing magnetic fields, emit curvature radiation which will generate $e^{\pm}$ pairs and subsequently an electromagnetic cascade. The uncertainties inherent to this scenario are significant, but it is very likely that one or more known pulsars contribute non-negligibly to the flux of cosmic ray electrons and positrons \cite{Gendelev} -- check Ref.~\cite{Latronico} for the implications of recent $\gamma$-ray observations of supernova remnants and pulsar wind nebulae on the positron fraction (and antiproton-to-proton ratio). A third hypothesis, put forward in Refs.~\cite{Blasi,BlasiSerpico}, posits that secondary particles are accelerated in the sites where hadronic primary cosmic rays are injected. If secondary particles (including electrons and positrons, but also antiprotons, boron and beryllium) are created in the acceleration site by spallation of primaries on the surrounding medium, then they will be accelerated themselves to high energies. A neat feature of this mechanism is that it can be easily tested since, besides a rising positron fraction, also rising secondary-to-primary ratios are predicted \cite{Sarkar}. Still other possibilities exist to accommodate the electron/positron excess, see e.g.~Section I of Ref.~\cite{Profumo} for an overview. In the following, we shall consider only the DM and the pulsar interpretations.

\par The main goal of the present work is to assess whether upcoming measurements of cosmic ray electrons and positrons will be sufficient to discriminate between different origins of the electron/positron excess. It is often claimed (see e.g. \cite{HallHooper}) that DM annihilations directly into $e^{+} e^{-}$ $-$ that produce a sharp spectral cutoff at the mass of the DM particle $-$ can be distinguished from a single pulsar spectrum with future data. 
Here, we start by assuming DM annihilations into leptons as the source of the cosmic ray lepton excess, anticipate upcoming measurements and evaluate to what degree one can discard the single pulsar hypothesis in that case. The inverse problem is also explored: assuming that the origin of the excess is a pulsar, and thus that a pulsar-like cut-off is detected, we quantify how well one can reject the DM hypothesis. Several DM masses are considered and, besides direct annihilation into electron-positron pairs, democratic annihilation into leptons (33\% $e^{+} e^{-}$, 33\% $\mu^{+} \mu^{-}$, 33\% $\tau^{+} \tau^{-}$) is also considered. In order to study the role of known pulsars, we make use of the ATNF catalogue \cite{atnf} and the $\gamma$-ray pulsars discovered by Fermi-LAT \cite{FermiCat,Fermi8}. Furthermore, the anisotropy potentially produced by individual pulsars is also discussed as a discriminating tool. 

\par We model the response of the Alpha Magnetic Spectrometer AMS-02 \cite{ams02}, scheduled to be installed in the International Space Station in 2011, including both systematic and statistical uncertainties. Future balloon-borne and ground-based experiments may also prove useful in measuring high-energy electrons and positrons, but we shall not consider them here since the associated systematics are likely larger than for AMS-02 (due to the influence of the atmosphere).

\par Several experimental results have been used along the work, namely measurements of the electron flux (CAPRICE \cite{caprice94}, HEAT \cite{heat9495e}, AMS-01 \cite{ams01}), the positron flux (CAPRICE \cite{caprice94}, HEAT \cite{heat9495e}, AMS-01 \cite{ams01}), the electron plus positron flux (HEAT \cite{heat9495e}, BETS \cite{bets01}, PPB-BETS \cite{PPBBETS}, ATIC \cite{ATIC}, H.E.S.S. \cite{hess08,hess09}, Fermi-LAT \cite{fermi,fermilat10}), and the positron fraction (CAPRICE \cite{caprice94}, HEAT \cite{heat9495,heat00}, AMS-01 \cite{ams01efrac}, PAMELA \cite{PAMELA10}).

\par The paper is organised as follows. In Sec.~\ref{secinj}, we outline the injection spectra of high-energy $e^{\pm}$ from DM annihilation and pulsars, and their propagation  through the galaxy. In Sec.~\ref{secexp}, we discuss our modelling of AMS-02 experimental capabilities.
In Sec.~\ref{DMPSR} we assess the capability of AMS-02 to distinguish between the DM and the pulsar hypotheses and, finally, in Sec.~\ref{secconc} we draw our conclusions.

\section{Injection and propagation of high-energy electrons and positrons}\label{secinj}

\par High-energy electrons and positrons in the galactic medium are mainly affected by two processes in their way to Earth: energy losses and diffusion. Energy losses in the multi-GeV range are dominated by inverse Compton scattering off cosmic microwave background, optical and infrared photons, and synchrotron emission. In the Thomson limit, these losses amount globally to $b(E)\simeq b_0 E^2$ with $b_0\sim 1.4 \times 10 ^{-16}$ GeV$^{-1}$s$^{-1}$. Diffusion instead is caused by the galactic magnetic irregularities and it is usually parametrized with a homogeneous power-law diffusion coefficient $D(E)=D_0(E/\textrm{GeV})^\delta$. Under these assumptions and neglecting convection and reacceleration, the number density of electrons and positrons per unit energy $n(\textbf{x},E,t)$ is driven by the transport equation
\begin{equation}\label{transpeq}
\frac{\partial n}{\partial t}= Q(\textbf{x},E,t) + D(E) \nabla^2 n + \frac{\partial}{\partial E} \left[ b(E) n \right] \quad ,
\end{equation}
$Q$ being the source term. Usually this equation is solved in steady state conditions ($\partial n / \partial t=0$) and inside a cylindrical diffusive halo of half-thickness $L$. The local interstellar flux then follows directly from the solution of equation \eqref{transpeq} through $\phi(\textbf{x}_{\odot},E)=\frac{v}{4\pi}n(\textbf{x}_{\odot},E)$, where $\textbf{x}_{\odot}$ is the position of the solar system in galactic coordinates.

\par In this section we briefly review the strategies adopted to compute the propagated injection spectra of electrons and positrons from each of the sources studied: DM annihilations and pulsar emission. On top of the flux generated by each source, one has of course to take into account the baseline astrophysical flux described in the previous section, where positrons are merely a by-product of the spallation of hadronic cosmic rays on the disk and electrons derive mainly from a galactic primary component. We shall refer to these standard yields as ``background'' since we are interested in the electron/positron excess. In order to be roughly compatible with Fermi-LAT measurements while not explaining the rising positron fraction observed by PAMELA \cite{PAMELA,PAMELA10}, we use ``model 1'' of Ref.~\cite{grasso} as our background reference setup. This model features a common high-energy injection index of 2.42 for electrons (above 4 GeV) and nuclei (above 9 GV), and the propagation parameters are fixed to the following values: $L=4$ kpc, $D_0=3.6\times 10 ^{28}$ cm$^2$/s and $\delta=0.33$. As in \cite{grasso}, we use GALPROP \cite{galpropsite, SM98} (version 50.1p) to compute the local fluxes of background electrons and positrons within the framework just described, and normalise to Fermi-LAT $e^{\pm}$ measurements at 100 GeV.

\par Before arriving at the top of the atmosphere, low-energy charged particles such as electrons and positrons are modulated by the solar wind. Our analysis focuses on energies above 10 GeV so that this effect is minimal. Nevertheless, all fluxes are modulated according to the force field approximation \cite{GA68} with a potential $\phi_F=550$ MV.

\par It is worth stressing that, even though different methods are used to compute the propagated fluxes of each source and background, the propagation parameters are kept fixed to the values mentioned above, and therefore our analysis is performed in a consistent propagation framework. However, as extensively discussed in the literature, the parameters $L$, $D_0$ and $\delta$ are not well-constrained by present cosmic-ray data, and hence the propagation configuration we use is not unique. Adopting other propagation parameters would affect differently the local fluxes of background, dark matter and pulsar electrons and positrons, but our results are representative of typical propagation models. Moreover, taking into account the uncertainties in the propagation parameters would add flexibility to fit the data and therefore strengthen our results. In the future it would be interesting to study how the AMS-02 ability to discriminate high-energy electron-positron sources depends on propagation details, perhaps using already the cosmic-ray nuclei measurements that will hopefully be provided by AMS-02 itself. Along these lines, another matter that deserves further investigation is the impact of the hardening of the proton and helium fluxes recently reported by CREAM \cite{creamhard} and ATIC \cite{atichard} at $\sim$ TeV/n energies (see also \cite{Lavallehard,Putzehard,DonatoSerpicohard}).

\subsection{Dark matter annihilations}

\par If dark matter is composed of annihilating Weakly Interacting Massive Particles (WIMPs), the galactic halo should be regarded as a continuous, nearby source of electrons and positrons. WIMPs of mass $m_{DM}$ and total averaged annihilation cross-section times velocity $\langle \sigma_{ann} v \rangle$ give rise to a constant source term
\begin{equation}\label{DMsource}
Q(\textbf{x},E)=\frac{1}{2}\left(\frac{\rho_{DM}(\textbf{x})}{m_{DM}}\right)^2 \langle \sigma_{ann} v \rangle \sum_k{BR_k \, \frac{dN^k}{dE}(E)} \quad ,
\end{equation}
where $\rho_{DM}$ is the density of dark matter, $k$ runs over the annihilation final states, $BR_k$ is the branching ratio of final state $k$ and $\frac{dN^k}{dE}$ is the electron or positron energy spectrum per annihilation into final state $k$. The factor 1/2 in equation \eqref{DMsource} is valid for Majorana self-annihilating fermions.

\par The distribution of dark matter in the Milky Way is poorly constrained by observations, but one can learn a great deal with the help of numerical simulations. In particular, one strong prediction of pure cold dark matter simulations is a large galactic population of virialised clumps in addition to a smooth halo component. Here, the DM density $\rho_{DM}(\textbf{x})$ is modelled according to the high-resolution dark matter only simulation Via Lactea II \cite{VL2}, where the smooth+clumpy distribution is well fitted by a Navarro-Frenk-White (NFW) profile and the abundance of subhalos of mass $M$ is proportional to $M^{-2}$. Within this setup, the spherically averaged local smooth DM density is $\rho_{\odot}=0.41$ GeV/cm$^3$ \cite{multi,cosmo}, and we extrapolate the results of the simulation down to a minimal subhalo mass of $10^{-6}$ M$_{\odot}$ which is a fiducial value for WIMPs \cite{Green} even though it may vary by several orders of magnitude \cite{Bringmann}. Further technical details can be found elsewhere \cite{multi,cosmo}.

\par Now, the transport equation \eqref{transpeq} with source term \eqref{DMsource} can be solved semi-analytically. We use the formulas derived in Refs.~\cite{Lavalle,Delahaye4} to compute the local flux of electrons and positrons created by DM annihilations in the smooth and clumpy components. Notice that the so-called substructure boost factor $-$ which quantifies the enhancement in the annihilation flux due to the presence of DM clumps with respect to the smooth only case $-$ is not a constant, but rather an energy- and particle-dependent function \cite{Lavalle}. Also, unlike sometimes assumed, it amounts to small rather than large enhancements.

\par We wish to point out that the distribution of DM particles in our Galaxy is uncertain, for what concerns the radial profile, the substructure population as well as the local DM density. However, such uncertainties are not crucial for our work and do not change our conclusions since different DM distributions or local densities would simply correspond to different normalisations $\langle \sigma_{ann} v \rangle$ and thus would produce essentially the same spectral features (unless the unlikely case of a nearby, massive clump is considered).

\par Two annihilation channels will be studied in the present paper: direct annihilation to $e^+ e^-$, and democratic annihilation to charged leptons (i.e.~$BR_{e^+e^-}=BR_{\mu^+\mu^-}=BR_{\tau^+\tau^-}=1/3$). While in the former case the injection spectrum of electrons and positrons is simply a Dirac delta at $E=m_{DM}$, the latter model injects $e^{\pm}$ pairs with a broad range of energies up to $m_{DM}$. Indeed, muons decay almost entirely to electrons and taus decay either to electrons, or to muons and hadronic particles that subsequently generate more electrons. In order to model these chains and find the $e^{\pm}$ injection spectra due to annihilations into $\mu^+\mu^-$ and $\tau^+\tau^-$, for each value of the DM mass we have generated one million annihilation events using PYTHIA \cite{Sjostrand:2000wi,Sjostrand:2006za,Sjostrand:2007gs} and used the results to reconstruct the electron/positron spectrum for a single annihilation. We have repeated this procedure for 26 values of the DM mass with constant linear spacing $\Delta m_{DM}=20\,$GeV between 100 and 600 GeV. Single annihilation spectra for intermediate values of the DM mass are obtained by interpolation. 

\par Other annihilation channels such as light quarks or gauge bosons are not considered because they lead to less characteristic features in the electron-positron spectrum and are thus more challenging to discriminate. Moreover, when normalized to the PAMELA positron fraction, models with annihilation to light quarks or gauge bosons usually produce an unacceptably large antiproton flux, at variance with observations \cite{cirelli,Delahaye2}.  Note that the annihilation modes under study, i.e.~direct annihilation to leptons only, are difficult to realise in the context of minimal supersymmetrical models, but are typical of the so-called \emph{leptophilic DM models} \cite{Fox:2008kb}, where tree-level DM annihilations to states other than leptons
are forbidden by an \emph{ad hoc} symmetry (see also \cite{Chun:2009zx} for a supersymmetric implementation of leptophilic DM).
Models of Kaluza-Klein (KK) DM also preferentially annihilate to charged leptons, in a democratic way (typically 20\% of time for each family), and despite the significant branching ratio to hadronic channels ($\simeq 35\%$), can satisfy the constraints on the antiproton flux   \cite{Hooper:2009fj}.  Thus, even if we do not refer to any particular particle physics model in the following, but rather choose a more phenomenological approach, our results concerning the ``democratic direct annihilation'' case can be considered as somewhat representative of what one would obtain in the context of leptophilic and KK DM models, once the differences in the relative branching ratios to leptons are taken into account. On the other hand, another class of models, inspired by the exciting DM scenario \cite{Cholis:2008vb,Cholis:2008qq}, are those in which a new, light force carrier is introduced  \cite{ArkaniHamed:2008qn}. In these models, the DM particle does not couple directly to Standard Model states, but instead annihilates to the new state, which in turn decays to Standard Model particles. If the new state is light enough, the decay to hadrons is kinematically forbidden. Thus the lack of hadrons is determined by the spectrum of the states and not by some kind of symmetry. However, the fact that the production of leptons happens through the annihilation to an intermediate state, makes the shape of the final $e^+/e^-$ spectrum qualitatively different from the case in which the DM annihilates directly to leptons (as in leptophilic and KK models). For this reason, care should be taken in interpreting our results in relation to this class of models. 

\par In this framework, for each annihilation model there are thus two free parameters in our analysis: the DM mass $m_{DM}$ and the annihilation cross-section $\langle \sigma_{ann} v \rangle$ (the latter entering basically as a normalisation factor). We shall use both in a model-independent, phenomenological manner in the remainder of the work.

\subsection{Pulsars}

\par Pulsars are highly magnetised, rotating neutron stars. Their main observational feature is the emission of pulsed and directional electromagnetic radiation (from radio photons to $\gamma$-rays) which suggests that the magnetic and rotation axes are misaligned. Unlike dark matter, pulsars are known sources of high-energy electrons and positrons, being thus a natural class of candidates to explain the cosmic ray lepton excess as pursued for example in Refs.~\cite{Hooper,Profumo,grasso}. In fact, the magnetosphere of a pulsar can easily host potential gaps in excess of $10^{12}$ V which accelerate primary electrons to TeV energies or above. These electrons quickly emit synchrotron and/or curvature radiation in the strong magnetic field, or upscatter background (radio, microwave, infrared or X-ray) photons to TeV-scale energies. Such $\gamma$-rays in turn are likely to interact with existing low energy photons or the magnetic fields and create high-energy $e^{\pm}$ pairs that subsequently generate an electromagnetic cascade. Further acceleration may also occur in the nebula or remnant surrounding the pulsar. Consequently, along with $\gamma$-rays (that have been detected), pulsars are very credible sources of high-energy cosmic ray electrons and positrons.

\par The rotation frequency of a pulsar $\Omega=2\pi/P$ decreases with time so that its rotational energy is dissipated at a rate $\dot{E}=\frac{d}{dt}\left(\frac{1}{2}I\Omega^2\right)=I\Omega \dot{\Omega}$, where $I=\frac{2}{5}M_{\star} R_{\star}^2$ is the moment of inertia of a spherical pulsar with mass $M_{\star}$ and radius $R_{\star}$. Magnetic braking, that accounts for the energy lost by magnetic dipole emission, is certainly a mechanism contributing to this spin-down behaviour even though there might be others. This process amounts to an energy loss rate \cite{Longair2} $\dot{E}_{mag}=-(8\pi\Omega^4 R_{\star}^6 B_{\star}^2)/(3 c^3 \mu_0)$, where $B_{\star}$ is the magnetic field at the star surface. In the simple case where magnetic braking dominates the pulsar spin-down luminosity, one can write $\dot{E}\simeq\dot{E}_{mag}$ that leads to 
\begin{equation}\label{freqeq}
\Omega(t)=\Omega_0\left(1+\frac{t}{\tau_0}\right)^{-1/2} \quad ,
\end{equation}
$\Omega_0$ being the initial ($t=0$) rotational frequency and $\tau_0=(3 I \mu_0 c^3)/(16 \pi R_{\star}^6 B_{\star}^2 \Omega_0^2)$. Notice that $\tau_0$ is essentially the luminosity decay time since $\dot{E}=I\Omega \dot{\Omega}\propto (1+t/\tau_0)^{-2}$. Therefore, very old pulsars ($t\gg \tau_0$) $-$ whose early $e^{\pm}$ emission has already diffused and diluted $-$ are not very likely to contribute to the bulk of the present local flux. Put another way, local high-energy electrons and positrons must have been produced recently as argued in Section \ref{secintro}. On the other hand, the electron-positron pairs produced by young pulsars need to diffuse through the pulsar nebula or the surrounding supernova remnant before reaching the ISM, which takes $t_0\sim10^4-10^5$ yr. The two opposite effects make mature pulsars $-$ of ages around $10^5$ yr $-$ the dominant source of a pulsar-induced $e^{\pm}$ galactic flux, and hence we shall focus on this type of pulsars in the present work. Let us notice at this point that the escape of $e^{\pm}$ pairs into the ISM is not fully understood yet -- an important step along this direction has been taken in \cite{BlasiAmato}.

\par The characteristic age of a pulsar is obtained by integrating $\dot{E}\simeq\dot{E}_{mag}$ under the assumption that the initial rotational frequency $\Omega_0$ is very large ($\Omega_0\gg\Omega$). This results in the well-known expression $t_{ch}=-\Omega/(2\dot{\Omega})=P/(2\dot{P})$. The actual age of the pulsar, $t_{PSR}$, may however differ from $t_{ch}$, the discrepancy being of order $10^4-10^5$ yr \cite{Profumo}. We shall disregard such discrepancy and identify $t_{PSR}$ with $t_{ch}$.

\par A major ingredient to compute the yield of electron-positron pairs injected by pulsars is the electronic energy output $E_{e^{\pm}}(t_{ch})=\eta_{e^{\pm}}\int_0^{t_{ch}}{dt \, |\dot{E}|}$, $\eta_{e^{\pm}}$ being the fraction of rotational energy transferred to electrons and positrons (we assume $\eta_{e^{+}}=\eta_{e^{-}}=\eta_{e^{\pm}}/2$). Using equation \eqref{freqeq} and assuming $t_{ch}\gg\tau_0$ (roughly valid for mature pulsars), one obtains
\begin{equation}\label{output}
E_{e^{\pm}}(t_{ch})\simeq \eta_{e^{\pm}} \frac{I}{2} \Omega_0^2 \simeq \eta_{e^{\pm}} |\dot{E}| \frac{t_{ch}^2}{\tau_0} \quad ,
\end{equation}
where in the last step we have used the approximate behaviour of equation \eqref{freqeq} for $t_{ch}\gg\tau_0$, $\Omega(t_{ch})\simeq \Omega_0 \left(t_{ch}/\tau_0\right)^{-1/2}$. We have assumed here that the output in electron-positron pairs is proportional to the rotational energy loss. This needs not be the case; for alternative scenarios see e.g.~Ref.~\cite{Profumo}.

\par As for the $e^{\pm}$ injection spectrum we use a rather phenomenological approach by considering a power-law with an exponential cutoff at high energies, $Q_{e^{\pm}}(E)=Q^0_{e^{\pm}} (E/\textrm{GeV})^{-\Gamma} \textrm{exp}(-E/E_{cut})$. Since electron-positron pairs and $\gamma$-rays are produced by the same physical process, the spectral index $\Gamma$ is probably correlated to the spectral indices seen in $\gamma$-ray observations of pulsars, or other multiwavelength measurements, which suggest $1 \lesssim \Gamma \lesssim 2$. The cutoff energy $E_{cut}$ instead is usually placed above the TeV. Finally, the normalisation $Q^0_{e^{\pm}}$ is fixed by the output \eqref{output} through $E_{e^{\pm}}(t_{ch})=\int_{m_e}^{\infty}{dE \, E \, Q_{e^{\pm}}(E)}$.

\par We now turn to the propagation of the injection spectrum just described. Pulsars may be modelled as bursting, point-like sources of $e^{\pm}$ pairs \cite{Buesching,Hooper,Profumo,Gendelev,grasso}. Indeed, the injection region is much smaller than the typical propagation distance covered by high-energy electrons and positrons, and the emission period is much shorter than the travel time to Earth. Thus, the source term reads 
\begin{equation}\label{PSRsource2}
Q_{e^{\pm}}(\textbf{x},E,t)= Q_{e^{\pm}}(E) \delta(\textbf{x}-\textbf{r}_0) \delta(t-t_0) \quad ,
\end{equation}
in which $\textbf{r}_0$ is the position of the pulsar and we shall consider times $t=t_{ch}\gg t_0$ as natural for mature pulsars. In addition, the problem assumes spherical symmetry if one considers local sources, namely at distances smaller than the half-thickness of the diffusive halo $L$. The effect of boundaries at $z=\pm L$ has been studied in \cite{Profumo} and shown to be negligible for the energies and pulsar ages of interest if $L> 1-2$ kpc. The spherically symmetric analytical solution of equation \eqref{transpeq} was derived in Ref.~\cite{Atoyan} for arbitrary energy losses and injection spectrum. Applying that result with the source term \eqref{PSRsource2}, the local density of electrons and positrons is found to be
\begin{equation}\label{ne1}
n_{e^{\pm}}(d,E,t_{ch})=\frac{Q_{e^{\pm}}(E') b(E') }{ b(E) \pi^{3/2} r_{dif}^3 (E,t_{ch}) } e^{-\frac{d^2}{r_{dif}^2(E,t_{ch})}} \, ,
\end{equation}
where $E'(E,\,t_{ch})=E/(1-E/E_{max}(t_{ch}))$ is the initial energy of a particle detected at energy $E$ assuming $b(E)=b_0 E^2$, $E_{max}(t_{ch})=1/(b_0 t_{ch})$ is the maximum energy allowed by losses, $d$ is the distance to the considered pulsar, and 
\begin{eqnarray*}
r_{dif}^2(E,t)&=& 4 \int_{E}^{E'}{d\tilde{E} \, D(\tilde{E})/b(\tilde{E})} \\ \nonumber
 &=& \frac{4 D(E) t E_{max}(t)}{(1-\delta) E} \left( 1-\left(1-\frac{E}{E_{max}(t)}\right)^{1-\delta} \right) \, .
\end{eqnarray*}

\par
\begin{figure}
 \centering
 \includegraphics[width=7.5cm,height=7.5cm]{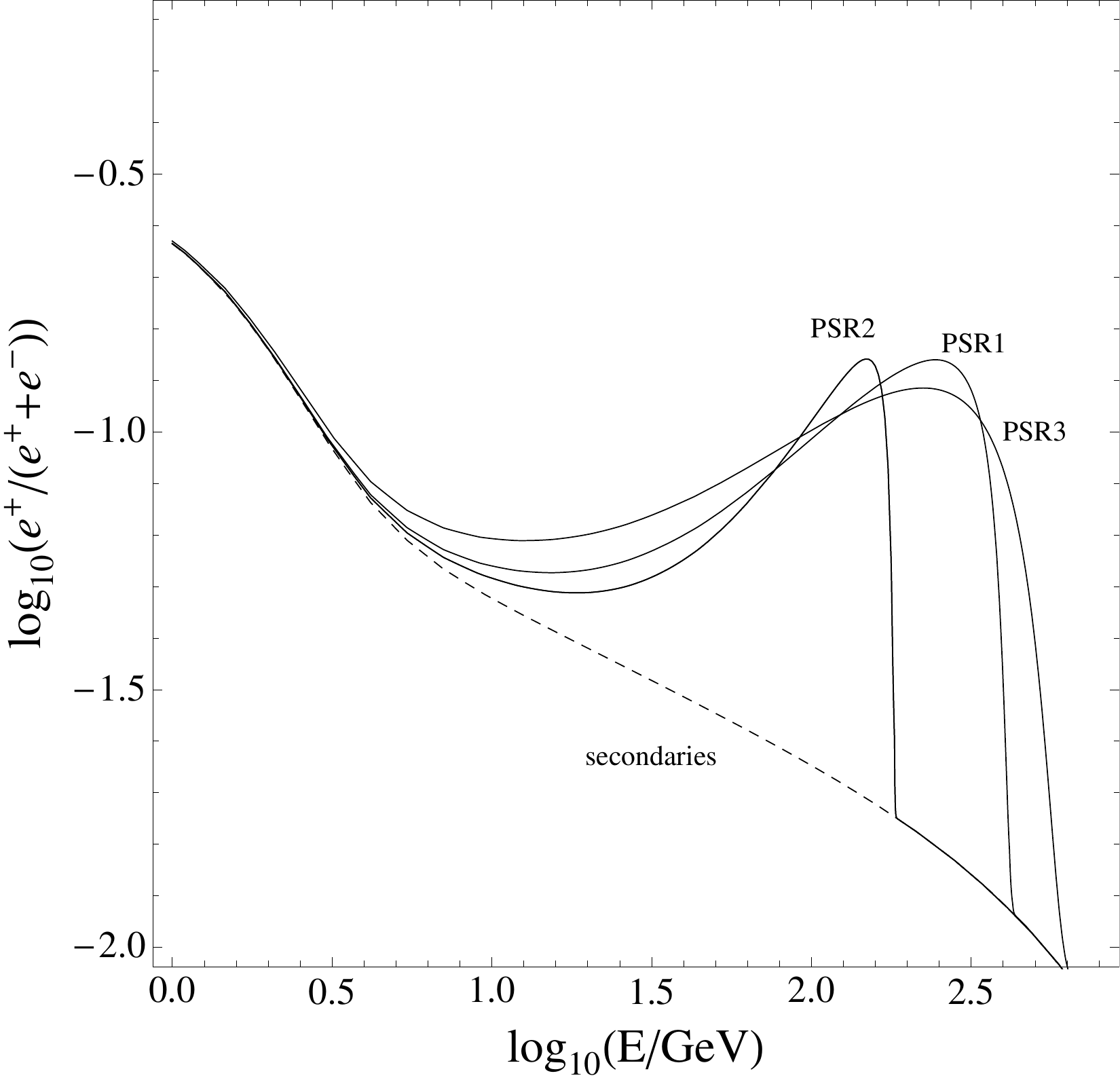}
 \caption{\fontsize{9}{9}\selectfont The positron fraction produced by the three pulsar fiducial models indicated in Table \ref{tab3}. The solid (dashed) line shows the total (background) contribution. In order of increasing cutoff energies, the plotted curves correspond to PSR2, PSR1 and PSR3.}\label{figposfracPSR}
\end{figure}

 Replacing $Q_{e^{\pm}}$ and $b$ in equation \eqref{ne1},
\begin{eqnarray}\nonumber
n_{e^{\pm}}(d,E,t_{ch})=\frac{Q^0_{e^{\pm}} (E/\textrm{GeV})^{-\Gamma} }{\pi^{3/2}r_{dif}^3 (E,t_{ch})} \left(1-\frac{E}{E_{max}(t_{ch})}\right)^{\Gamma-2} \\ \label{ne2}
\times \, \textrm{exp} \left(-\frac{E/E_{cut}}{1-E/E_{max}(t_{ch})} - \frac{d^2}{r_{dif}^2(E,t_{ch})}\right) 
\end{eqnarray}
for $E<E_{max}$, and 0 otherwise. Notice that $n_{e^{-}}=n_{e^{+}}=n_{e^{\pm}}/2$ since we are assuming $\eta_{e^{+}}=\eta_{e^{-}}=\eta_{e^{\pm}}/2$. It is worth pointing out that the spectrum in equation \eqref{ne2} features a cutoff at $\textrm{min}(E_{max},E_{cut})$. The nature and morphology of the cutoff is different in the two cases, $E_{max} \gtrless E_{cut}$ . For instance, for very large cutoff energies $E_{cut}$, the maximum energy local electrons and positrons can reach is fixed by energy losses and is therefore a function of the pulsar age. To illustrate this point we pick the three fiducial pulsar setups in Table \ref{tab3} that yield the positron fractions shown in Figure \ref{figposfracPSR}. Notice the different cutoff steepness in the each case -- we shall return to this issue later on.

\begin{table*}
\centering
\fontsize{9}{9}\selectfont
\begin{tabular}{c|ccccccc}
\hline
\hline
      & $\Gamma$ & $E_{cut}$ $[\textrm{GeV}]$ & $|\dot{E}|$ $[\textrm{erg/s}]$ & $t_{ch}$ $[\textrm{yr}]$ & $E_{max}$ $[\textrm{GeV}]$ & $d$ $[\textrm{kpc}]$  & $f$ \\
\hline
PSR1 & 1.7 & 1000 & $10^{35}$ & $5\times 10^5$ & 453 & 0.75 & 0.9 \\
PSR2 & 1.3 & 1000 & $4\times10^{34}$ & $12\times 10^5$ & 189 & 0.4 & 0.9 \\
PSR3 & 1.7 & 500 & $8\times10^{34}$ & $3\times 10^5$ & 755 & 0.2 & 0.9 \\
\hline
\end{tabular}
\caption{\fontsize{9}{9}\selectfont The pulsar fiducial setups. In all cases $\tau_0=10^4$ yr and $\eta_{e^{\pm}}=0.4$ as explained in the text. $f$ represents the background rescaling factor (see Section \ref{DMPSR}).}\label{tab3}
\end{table*}

\par Despite the fact that diffusion erases the initial direction of charged particles, the output of a pulsar still creates a dipole anisotropy on the cosmic ray electron-positron sky. Given a certain direction in the sky, let $N_{max}$ denote the number of detected photons in the hemisphere centred in that direction, and $N_{min}$ the number of photons in the opposite hemisphere. The size of the anisotropy is energy-dependent and given by $\delta_{e^{\pm}}=(N_{max}-N_{min})/(N_{max}+N_{min})=3 D(E) |\vec{\nabla} n_{e^{\pm}}(E)|/(cn_{e^{\pm}}^{tot})$, $n_{e^{\pm}}^{tot}$ being the sum of the pulsar signal $n_{e^{\pm}}$ and the background. Applying equation \eqref{ne2} one readily obtains
\begin{equation}\label{anis}
\delta_{e^{\pm}}(d,E,t_{ch})=\frac{6 D(E) d}{c \, r_{dif}^2(E,t_{ch})} \frac{n_{e^{\pm}}(d,E,t_{ch})}{n_{e^{\pm}}^{tot}(d,E,t_{ch})} \quad .
\end{equation}
The expressions for the electron- or positron-only anisotropies are analogous.

\par In order to explore the possible features of the electron-positron spectrum generated by mature pulsars we scan the parameter space defined by the spin-down luminosity $|\dot{E}|$, the distance $d$ and the characteristic age $t_{ch}$ in the ranges $|\dot{E}|=10^{32}-10^{36}$ erg/s, $d=0.1-5$ kpc and $t_{ch}=10^4-10^7$ yr. Following \cite{Profumo,Gendelev}, we adopt $\tau_0=10^4$ yr and the rather large $e^{\pm}$ fractional output $\eta_{e^{\pm}}=0.4$. Notice however that $\tau_0$ and $\eta_{e^{\pm}}$ are both degenerate with $|\dot{E}|$ for mature pulsars $-$ check equation \eqref{output} $-$ so that our results may be simply rescaled to account for different values of $\eta_{e^{\pm}}$ for instance. For the spectral injection index and the cutoff energy we take the reference values $\Gamma=1.7$ and $E_{cut}=1$ TeV, but we also address the effect of varying these parameters in the ranges $1.3 \leq \Gamma \leq 1.9$ and $0.5 \leq E_{cut} \leq 10$ TeV. Once $|\dot{E}|$, $d$, $t_{ch}$, $\eta_{e^{\pm}}$, $\tau_0$, $\Gamma$ and $E_{cut}$ are specified, the local flux of high-energy electrons and positrons produced by a single mature pulsar is unambiguously fixed through equation \eqref{ne2}.

\par Throughout the work we shall use the ATNF catalogue \cite{atnf}, that contains the most comprehensive list of pulsars observed in different wavelengths, as well as Fermi-LAT $\gamma$-ray pulsars including both the objects listed in the catalogue \cite{FermiCat} (for the pulsars in the catalogue without distance estimate we use the $\gamma$-ray distance as determined by \cite{Gendelev}) and eight recently discovered ones \cite{Fermi8}. This will give us insight on the regions in parameter space occupied by known pulsars and their role in explaining the electron/positron excess. Every pulsar catalogue suffers from more or less important uncertainties, in particular related to the distance and characteristic age estimates. Hence, we will show the distance error in Fermi-LAT pulsars and the age uncertainty due to the timing measurements reported in ATNF catalogue.

\section{Modelling AMS-02 experimental capabilities}\label{secexp}

\par In order to study the prospects for discriminating the source of the electronic component in cosmic rays, we focus on the expected performances of the Alpha Magnetic Spectrometer (AMS-02) \cite{ams02}. Being a large-acceptance spectrometer in space, AMS-02 will likely be invaluable in the measurement of cosmic ray nuclei spectra and ratios. Moreover, good rejection capabilities will enable a precise determination of cosmic ray electrons and positrons in the GeV$-$TeV range. We start by studying the energy range 1$-$300 GeV for both electrons and positrons \cite{SpadaPlanck}, where the energy resolution has been determined using a test beam at CERN: $\frac{\Delta E}{E}=\sqrt{(0.106/\sqrt{E/\textrm{GeV}})^2+(0.0125)^2}$ \cite{Kounine}. This corresponds to $\sim$ 10.7\% (1.4\%) energy resolution at $E=1\,(300)$ GeV, and conservatively we assume 15 energy bins per decade. The recent replacement of the superconducting magnet by the permanent one will allow AMS-02 to last more than the initial mission duration of 3 years, even though with a downgraded performance. In this work we take a data-taking period of 18 years. We have, however, verified that, for an 1 year period, the projected AMS-02 data shown in the following is essentially unchanged since the uncertainties are dominated by systematics.

\par For a given energy bin of central energy $E_b$ and width $\delta E$, we estimate the detected number of X particles as 
\begin{equation}\label{NX}
N_X=\Delta t \, \delta E \, A_X \int{dE' \, \phi_X(E') \, \frac{e^{-\frac{(E'-E_b)^2}{2 \sigma^2}}}{\sqrt{2\pi \sigma^2}} } \quad ,
\end{equation}
where $\sigma=\Delta E (E_b)/2$, $\Delta t$ is the operating time (taken to be 18 year as stated in the last paragraph), $A_X$ is the geometrical acceptance of the instrument for X particles and $\phi_X$ is the differential flux. Following \cite{maestro}, we take a mean acceptance for electrons and positrons of $A_{e^{-}}=A_{e^{+}}=0.045$ m$^2$sr, valid in the energy range 1$-$300 GeV. Notice that the Gaussian smearing in the above formula is of particular importance for spectral features such as cutoffs, while being less relevant for smooth parts of the energy spectrum.

\par The relative statistical uncertainty inherent to the measurement of the flux $\phi_X$ in a certain energy bin is simply $1/\sqrt{N_X}$. Systematic errors instead result mainly from the misidentification of other particles as electrons and/or positrons. Protons in particular represent the major background for the measurement of the electronic component. To estimate their number in each energy bin we adopt the proton flux measured by BESS \cite{bessp} (extrapolated where needed) and use the expression in the last paragraph without Gaussian smearing and with $A_p=0.3$ m$^2$sr. For electrons, we adopt a rejection power against protons $e^{-}:p=3\times 10^5$ \cite{Schmanau,Casaus2009}. For positrons, the rejection powers against protons and electrons are respectively $e^{+}:p=3\times 10^5$ and $e^{+}:e^{-}=10^4$ \cite{maestro}. In a given energy bin, the relative systematic uncertainties are thus $\frac{N_p/N_{e^{-}}}{e^{-}:p}$ for the electron flux, and $\frac{N_p/N_{e^{+}}}{e^{+}:p}+\frac{N_{e^{-}}/N_{e^{+}}}{e^{+}:e^{-}}$  for the positron flux. For the rest of this work we shall add in quadrature systematic and statistical uncertainties.

\par It is not entirely clear to what extent will AMS-02 measure electrons and positrons above a few hundred GeV. In addition to the range 1$-$300 GeV, we also consider the window 300$-$800 GeV with the characteristics detailed above. Note that this is an optimistic approach since the AMS-02 performance will be likely worse above a few hundred GeV.

\par Finally, we analyse the prospects for detecting a dipole anisotropy in the flux of cosmic ray electrons and positrons. A nearby source may in fact produce a non-negligible anisotropy, in particular at high energies. Experimentally, the anisotropy measurement is limited by the presence of the (nearly) isotropic electron-positron background. Suppose $N \pm \Delta N$ is the number of background particles (electrons, positrons or both) in a hemisphere along a certain direction. Here, $\Delta N$ represents a global uncertainty including systematic and statistical errors. Then, the minimum detectable anisotropy at $n$ sigma is simply $\delta_{0,n\sigma}=n \Delta N / N$. Later on we shall use the 2$\sigma$ positron and electron plus positron anisotropy reaches of AMS-02 in 18 years, modelled according to the above formula and the details given in the previous paragraphs. We will also apply the 2$\sigma$ electron plus positron anisotropy reach of Fermi-LAT ($A_{e^{\pm}}\sim 1$ m$^2$sr) after 5 years and considering statistical errors only, even though this is a rather optimistic limit. A dedicated search for $e^{\pm}$ anisotropies was already performed by the Fermi-LAT collaboration with 1 year of data \cite{fermianis} producing interesting upper limits on $\delta_{e^{\pm}}$.

\section{Distinguishing dark matter and single pulsar spectra}\label{DMPSR}

\begin{figure}
 \centering
 \includegraphics[width=7.5cm,height=7.5cm]{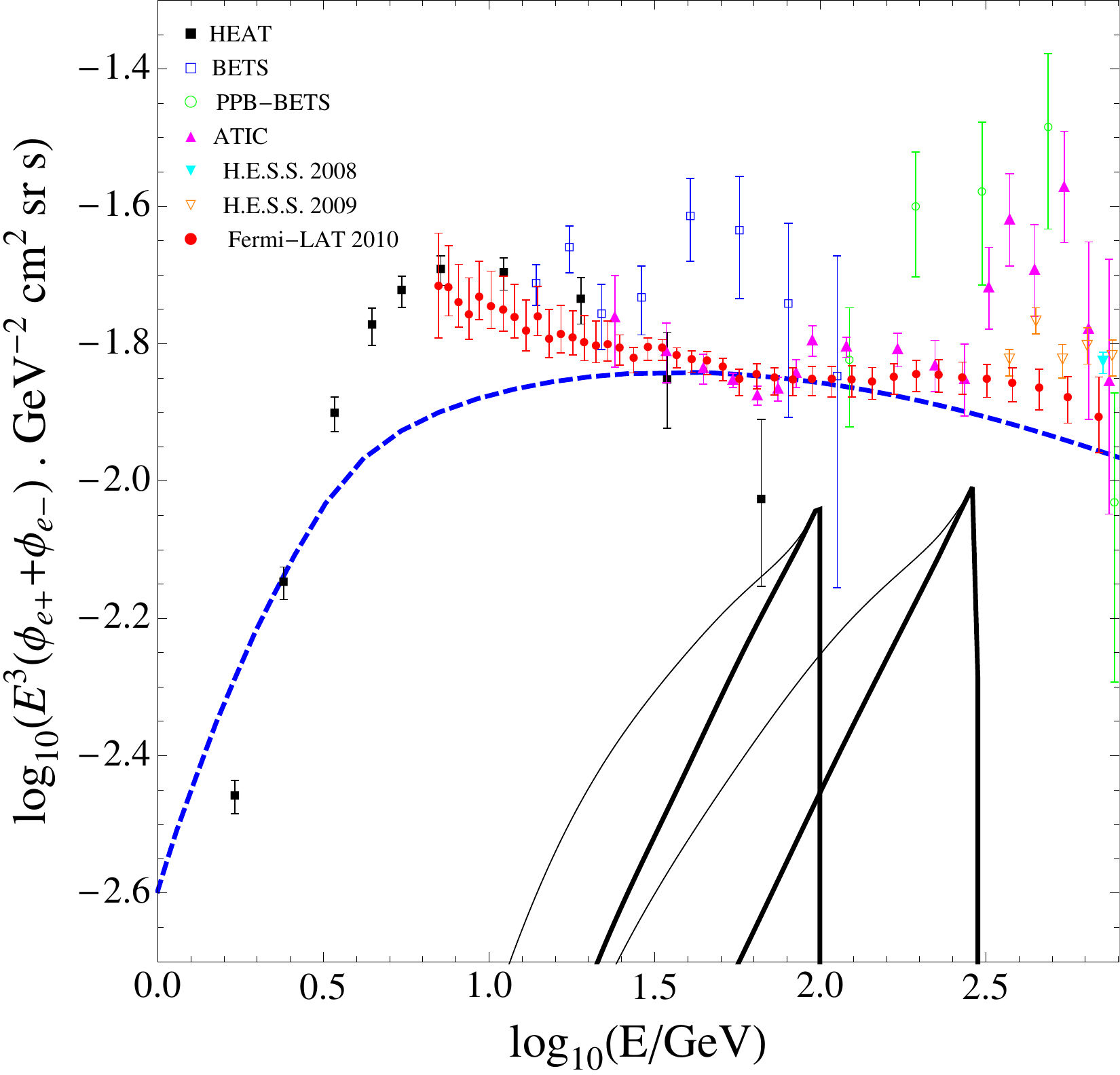}
 \caption{\fontsize{9}{9}\selectfont The electron plus positron spectrum produced by DM annihilations into $e^{+}e^{-}$ (thick solid lines) and democratic leptons (thin solid lines). DM masses are fixed to 100 and 300 GeV, and annihilation cross-sections are chosen arbitrarily for plotting purposes. The dashed line indicates the background used in this work, corresponding to  ``model 1'' of Ref.~\cite{grasso}. The data sets include HEAT \cite{heat9495e}, BETS \cite{bets01}, PPB-BETS \cite{PPBBETS}, ATIC \cite{ATIC}, H.E.S.S.~2008 \cite{hess08}, H.E.S.S.~2009 \cite{hess09} and Fermi-LAT \cite{fermilat10}.}\label{figcutoffs}
\end{figure}

\begin{table}
\centering
\fontsize{9}{9}\selectfont
\begin{tabular}{c|ccc}
\hline
\hline
      & $m_{DM}$ $[\textrm{GeV}]$ & $\langle \sigma_{ann} v \rangle$ $[\textrm{cm}^3/s]$ & $f$ \\
\hline
DM1 & 100 & $5.0\times10^{-26}$ & 0.97 \\
DM2 & 300 & $3.5\times10^{-25}$ & 0.87 \\
DM3 & 500 & $9.0\times10^{-25}$ & 0.83 \\
\hline
\end{tabular}
\caption{\fontsize{9}{9}\selectfont The dark matter fiducial setups. Besides mass $m_{DM}$ and annihilation cross-section $\langle \sigma_{ann} v \rangle$, also shown is $f$, the background rescaling factor. In all cases direct annihilation into $e^+e^-$ is assumed.}\label{tab1}
\end{table}

\par In this section we quantitatively study the impact of future cosmic-ray electron data on distinguishing dark matter and single pulsars as the source of the lepton excess. We start by assuming a dark matter origin and assess to what extent one can discard the pulsar hypothesis. If dark matter particles directly annihilate into electron-positron pairs only or democratically into the three charged leptons, then a rather abrupt cutoff at the DM mass is expected. This is illustrated in Figure \ref{figcutoffs}, where the expected $e^{\pm}$ background (detailed in Section \ref{secinj}) is also shown. In the following we shall focus on direct annihilations into $e^{\pm}$ since its extreme spectral feature is in principle more difficult to mimic with pulsars. We consider three phenomenological sets of DM properties $-$ summarised in Table \ref{tab1} $-$ that feature a rise in the positron fraction as seen by PAMELA and a cutoff at 100, 300 and 500 GeV. The normalisation of the baseline $e^{\pm}$ flux was rescaled by a factor $f$ (with respect to Fermi-LAT data point at 100 GeV) in order to give some room for the extra component. All three models produce electron plus positron fluxes compatible with Fermi-LAT and H.E.S.S. at the 3$\sigma$ level. We use these three models as our fiducial models.

\begin{figure*}
 \centering
 \includegraphics[width=0.32\textwidth]{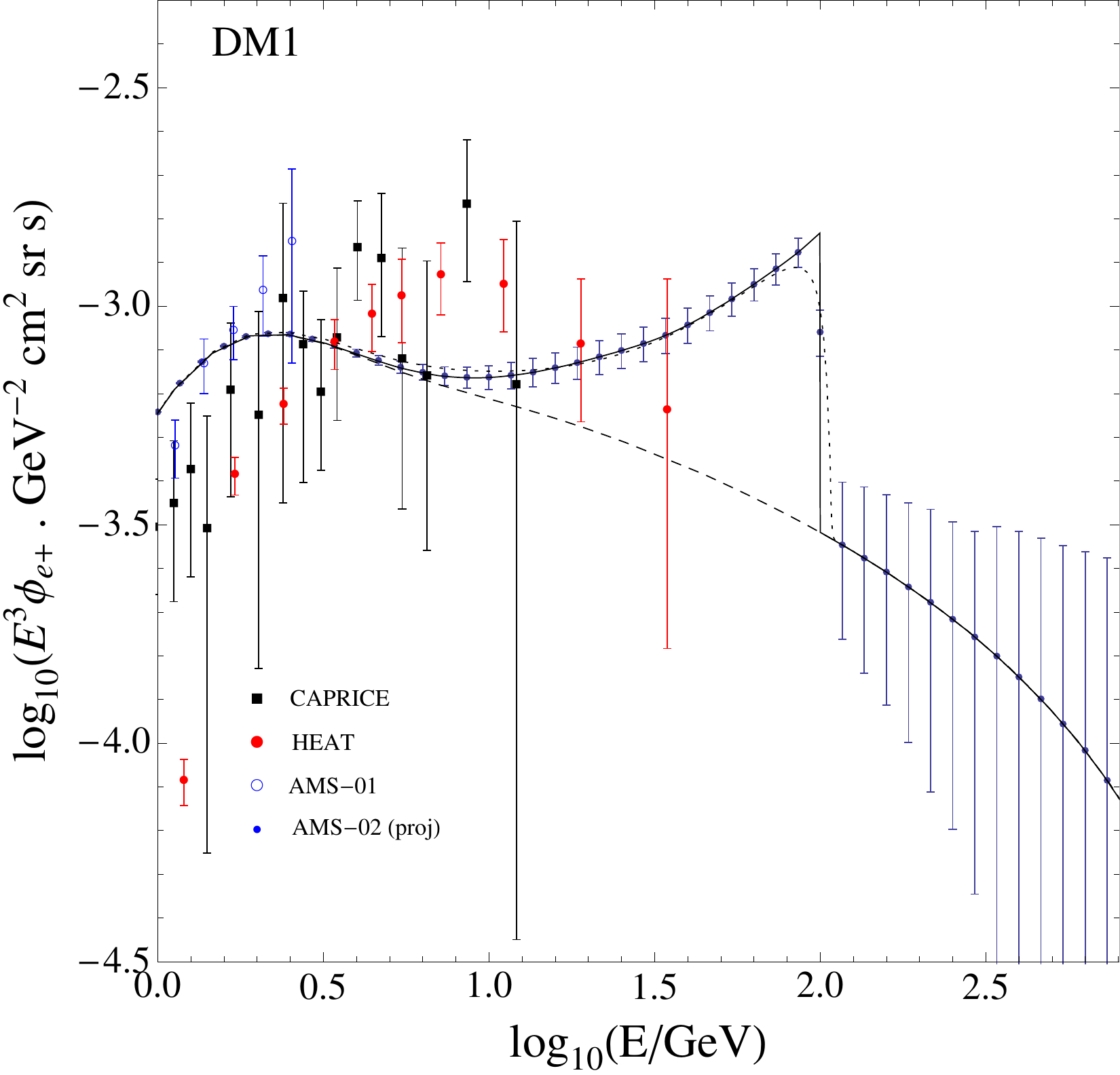}
 \includegraphics[width=0.32\textwidth]{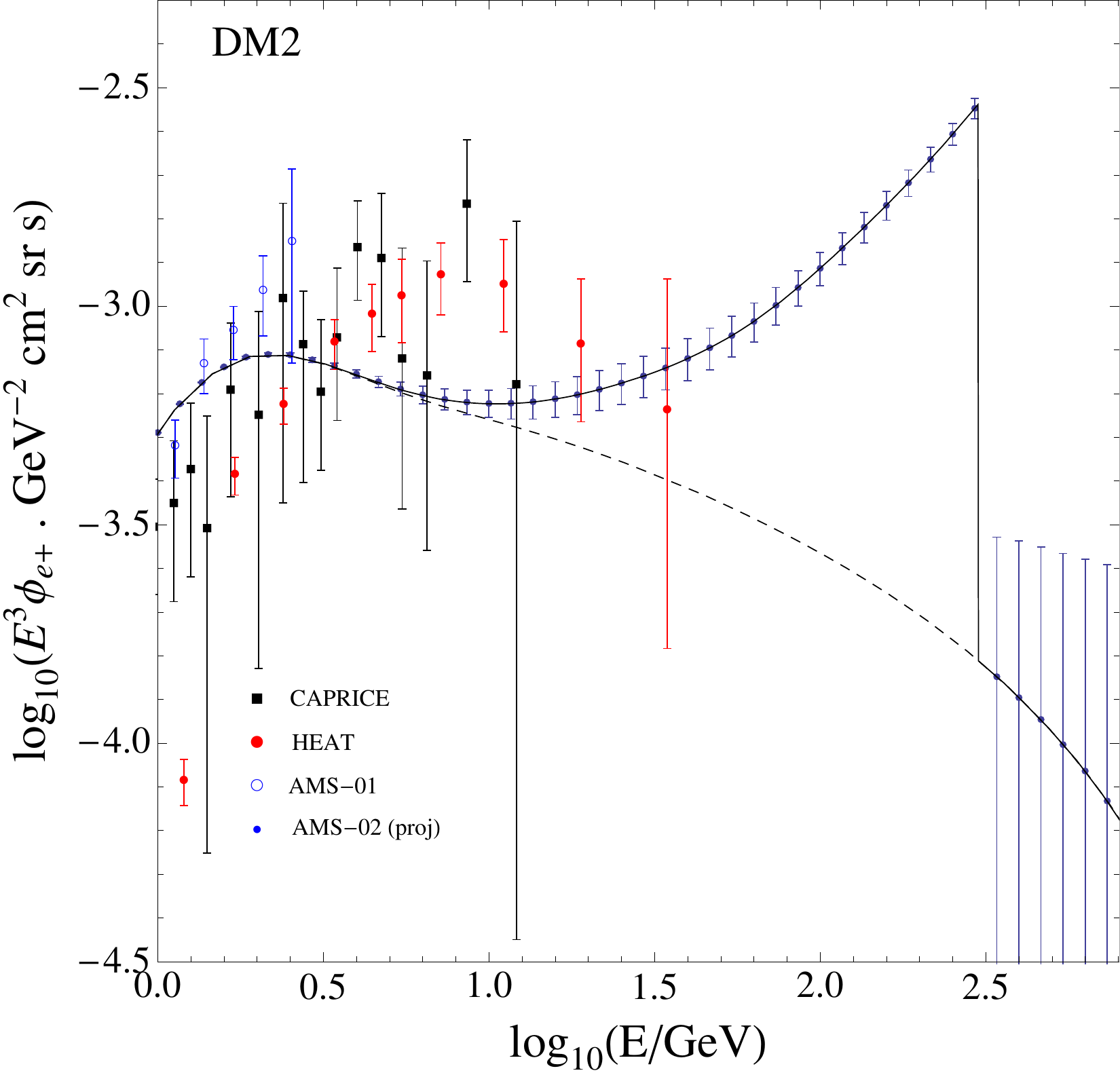}
 \includegraphics[width=0.32\textwidth]{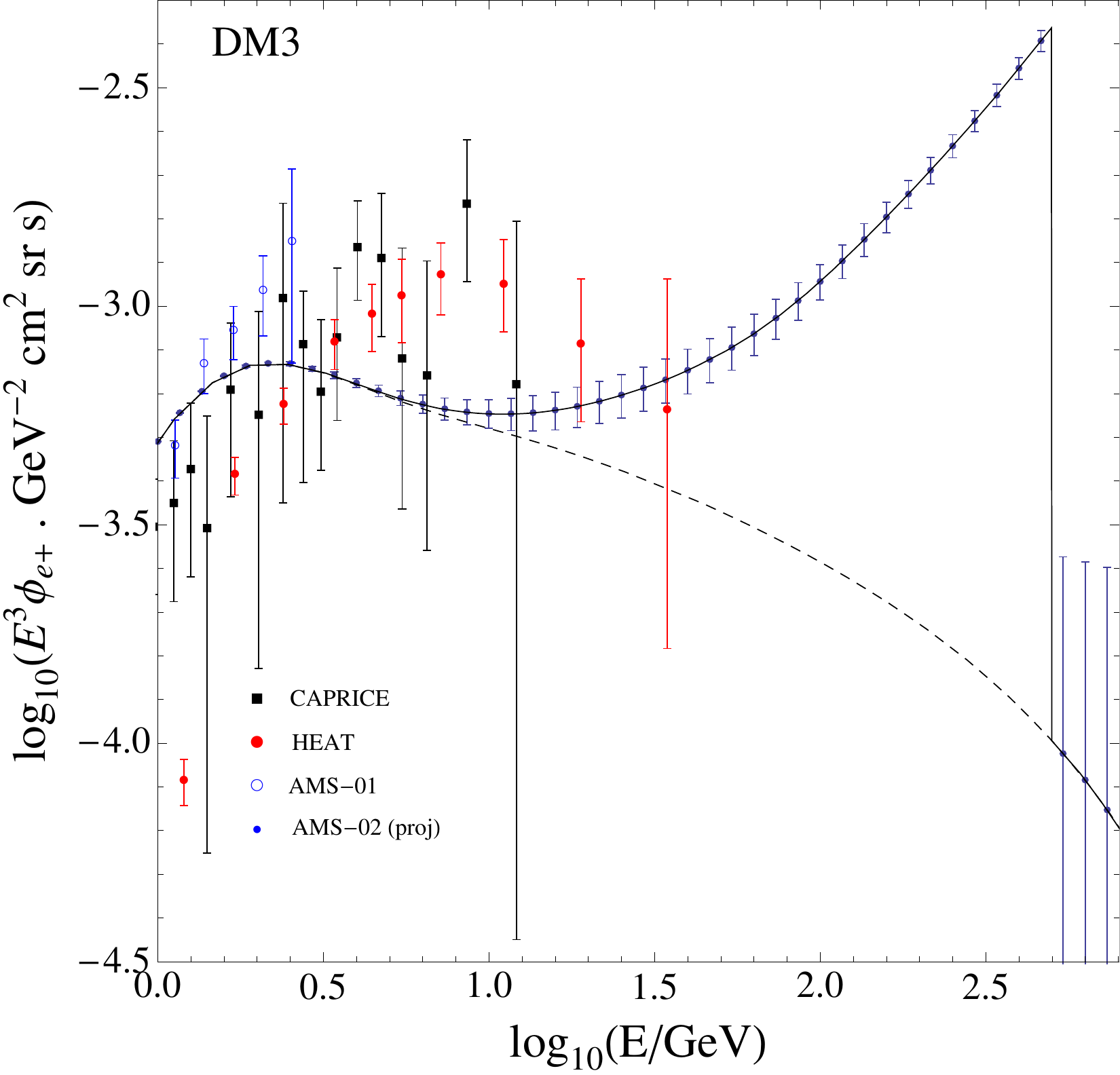}\\
 \includegraphics[width=0.32\textwidth]{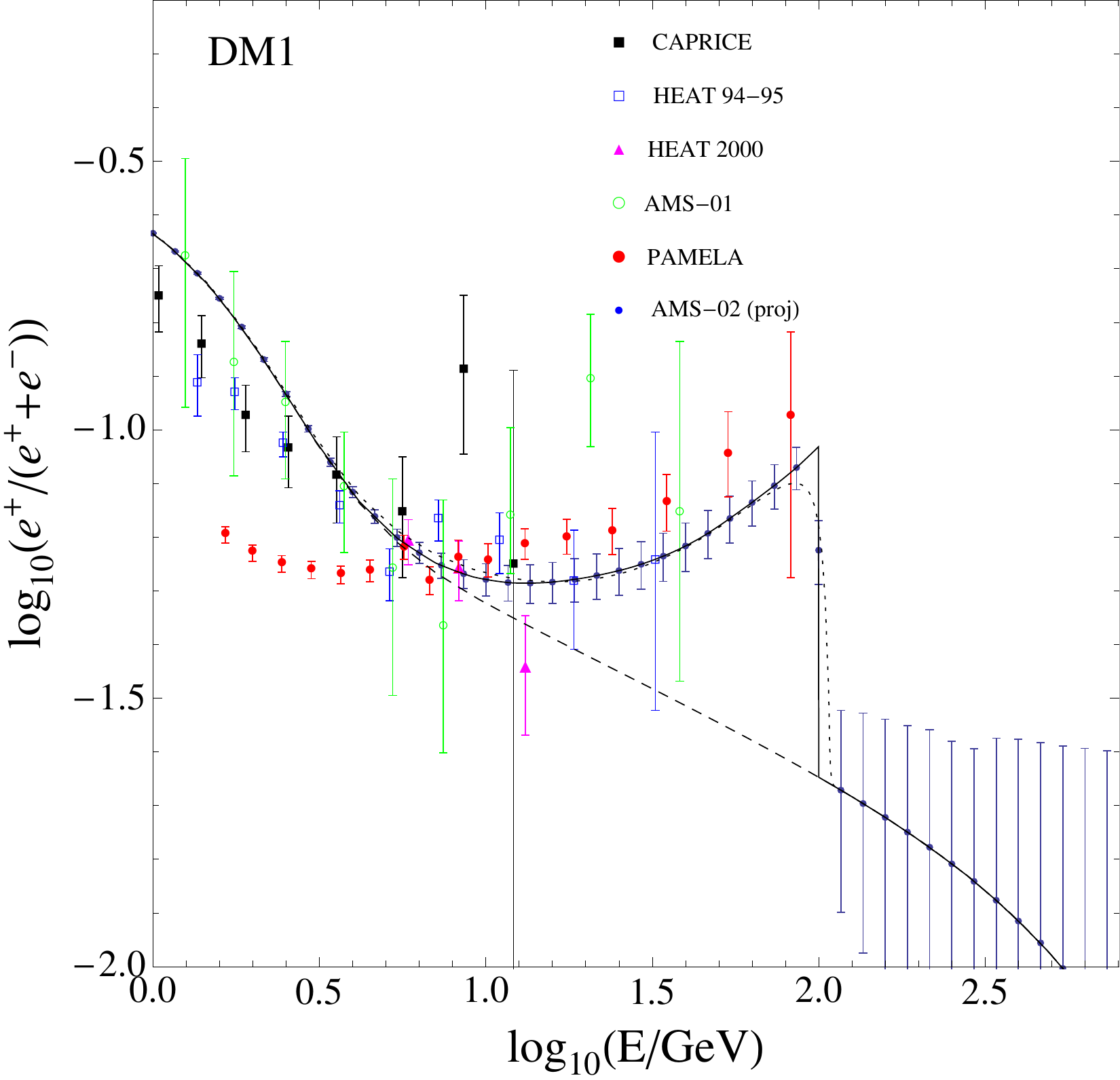}
 \includegraphics[width=0.32\textwidth]{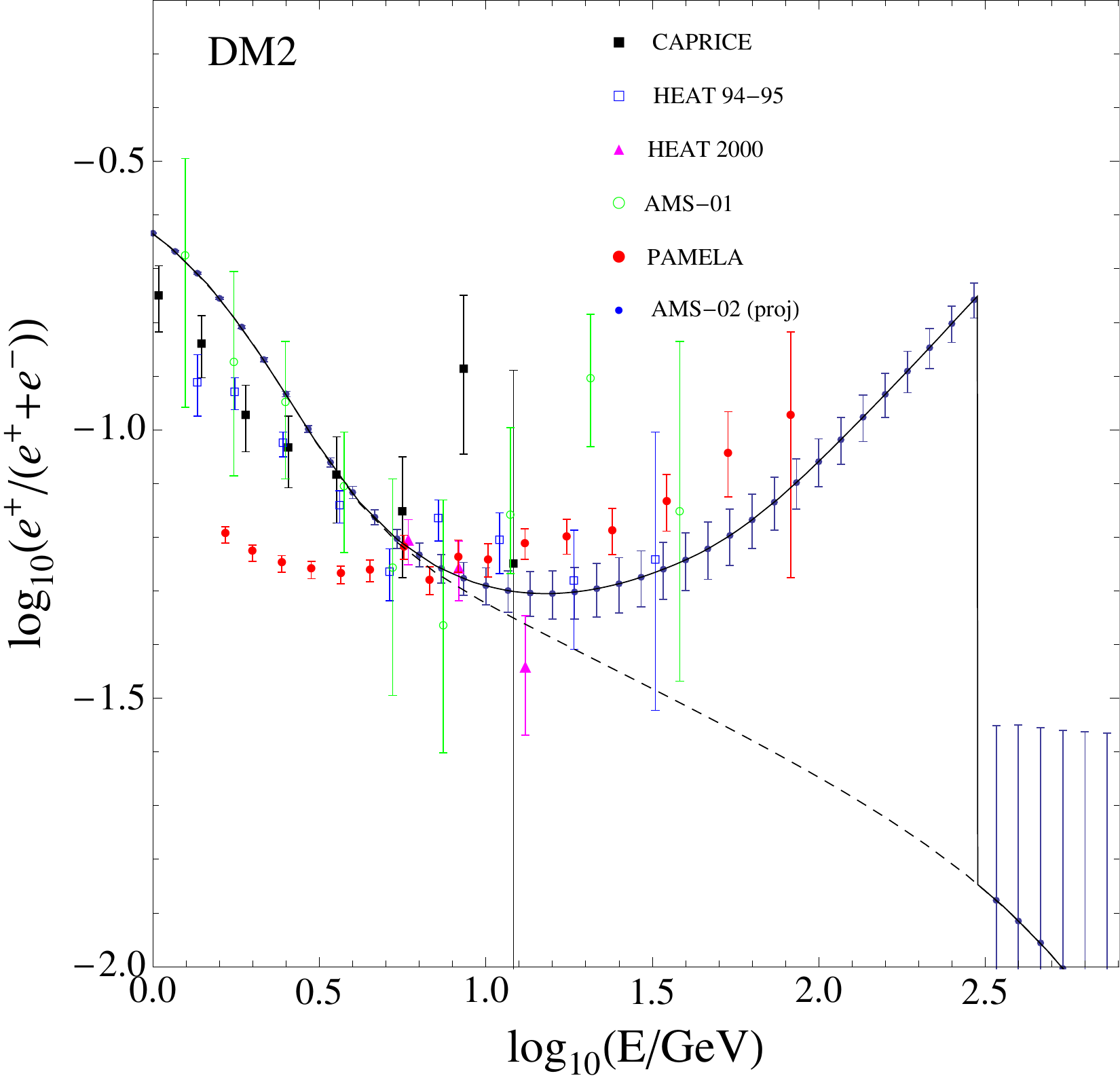}
 \includegraphics[width=0.32\textwidth]{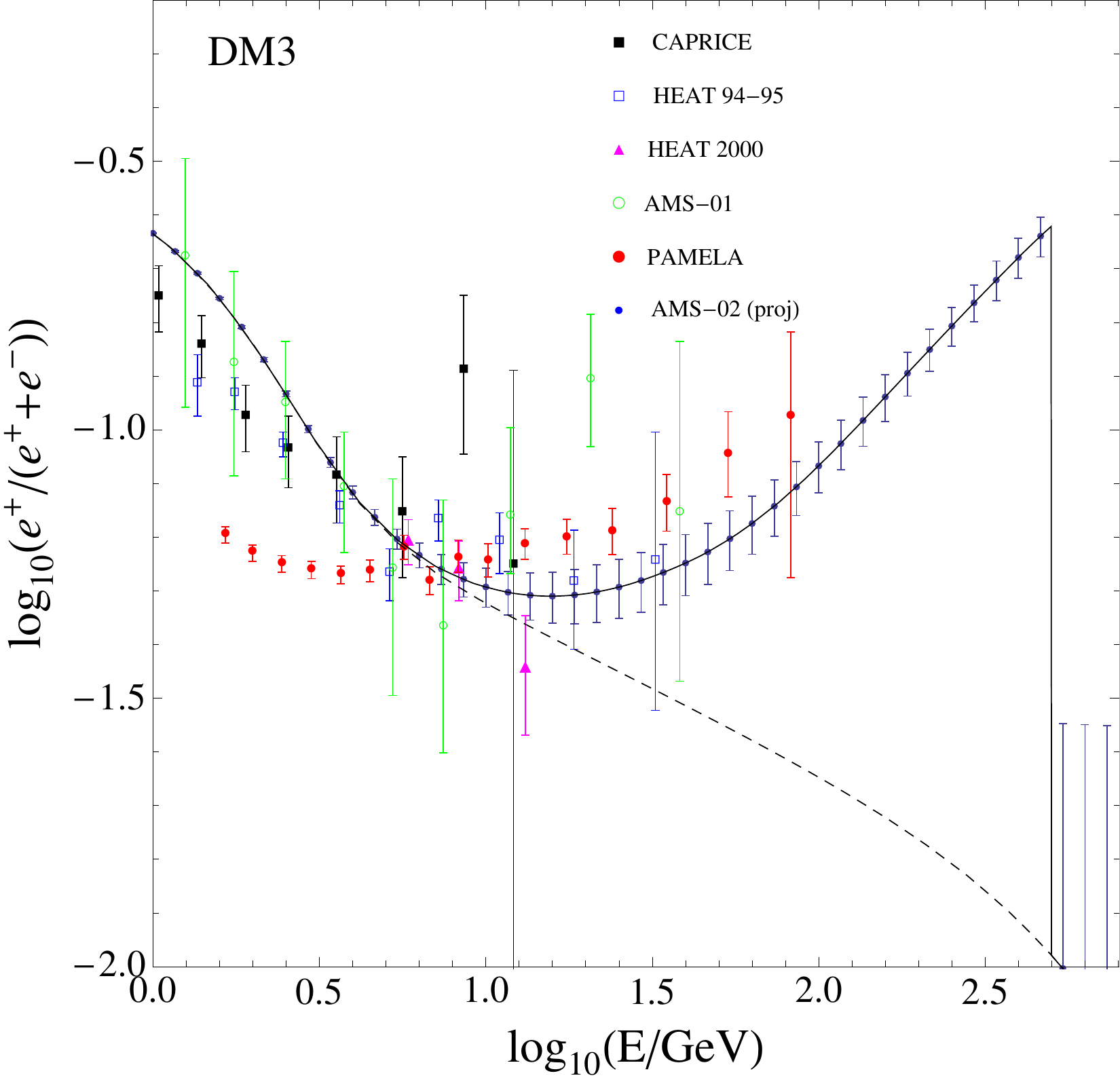}
 \caption{\fontsize{9}{9}\selectfont Mock AMS-02 data (filled blue circles) and corresponding error bars for the DM fiducial setups in Table \ref{tab1}. The solid (dashed) line shows the total (background) contribution. The top row of frames features the positron flux while the bottom one the positron fraction. From left to right, the columns correspond to DM1, DM2 and DM3. In the left column the dotted line shows a pulsar spectrum that fits well the AMS-02 mock data. For the positron flux, the data sets include CAPRICE \cite{caprice94}, HEAT \cite{heat9495e}, AMS-01 \cite{ams01}. For the positron fraction, the data sets include CAPRICE \cite{caprice94}, HEAT 1994-1995 \cite{heat9495}, HEAT 2000 \cite{heat00}, AMS-01 \cite{ams01efrac} and PAMELA \cite{PAMELA10}.}\label{figDMdata}
\end{figure*}

\begin{figure*}
 \centering
\hspace{-1.cm}
 \includegraphics[width=9.cm,height=7.5cm]{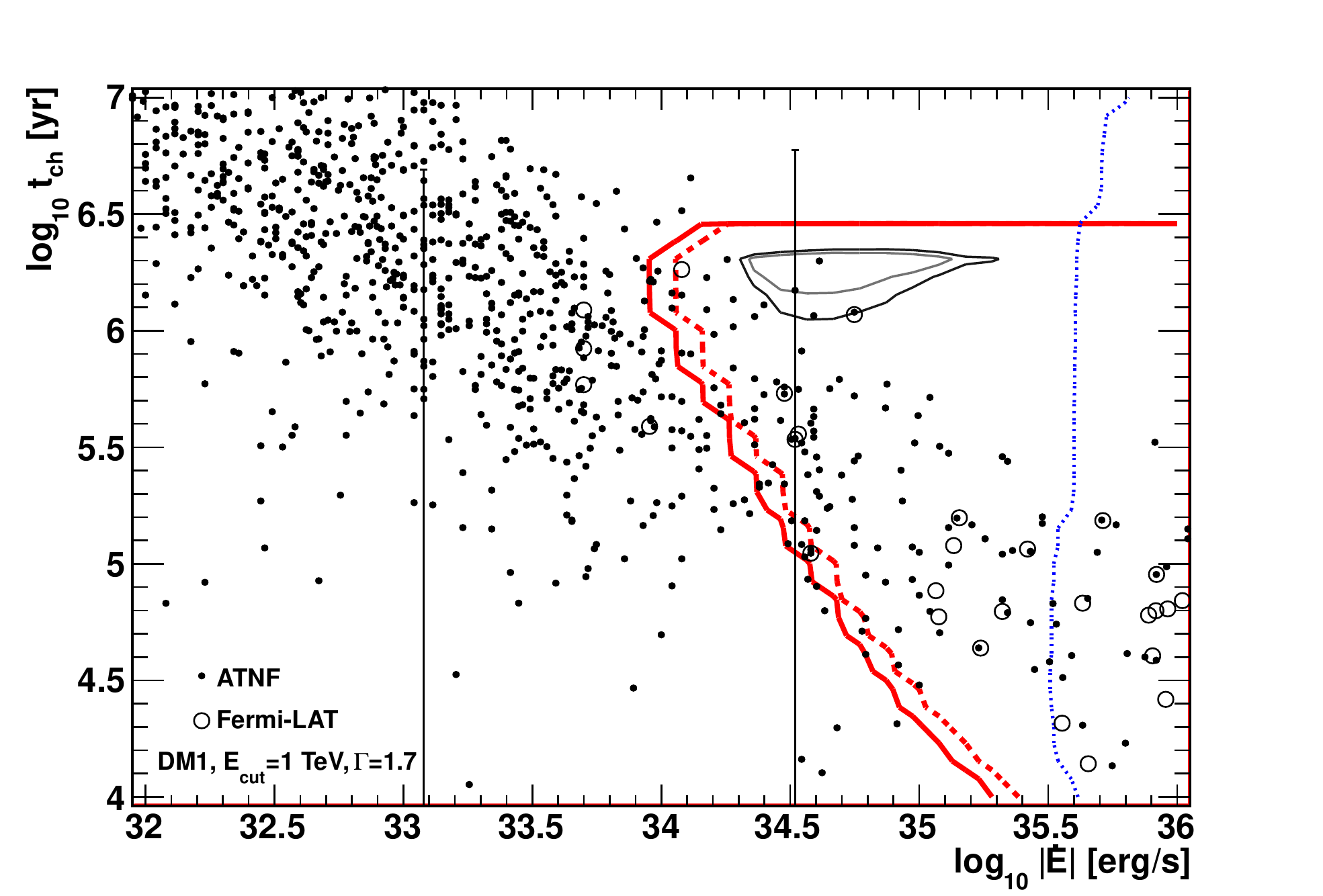}
 \includegraphics[width=9.cm,height=7.5cm]{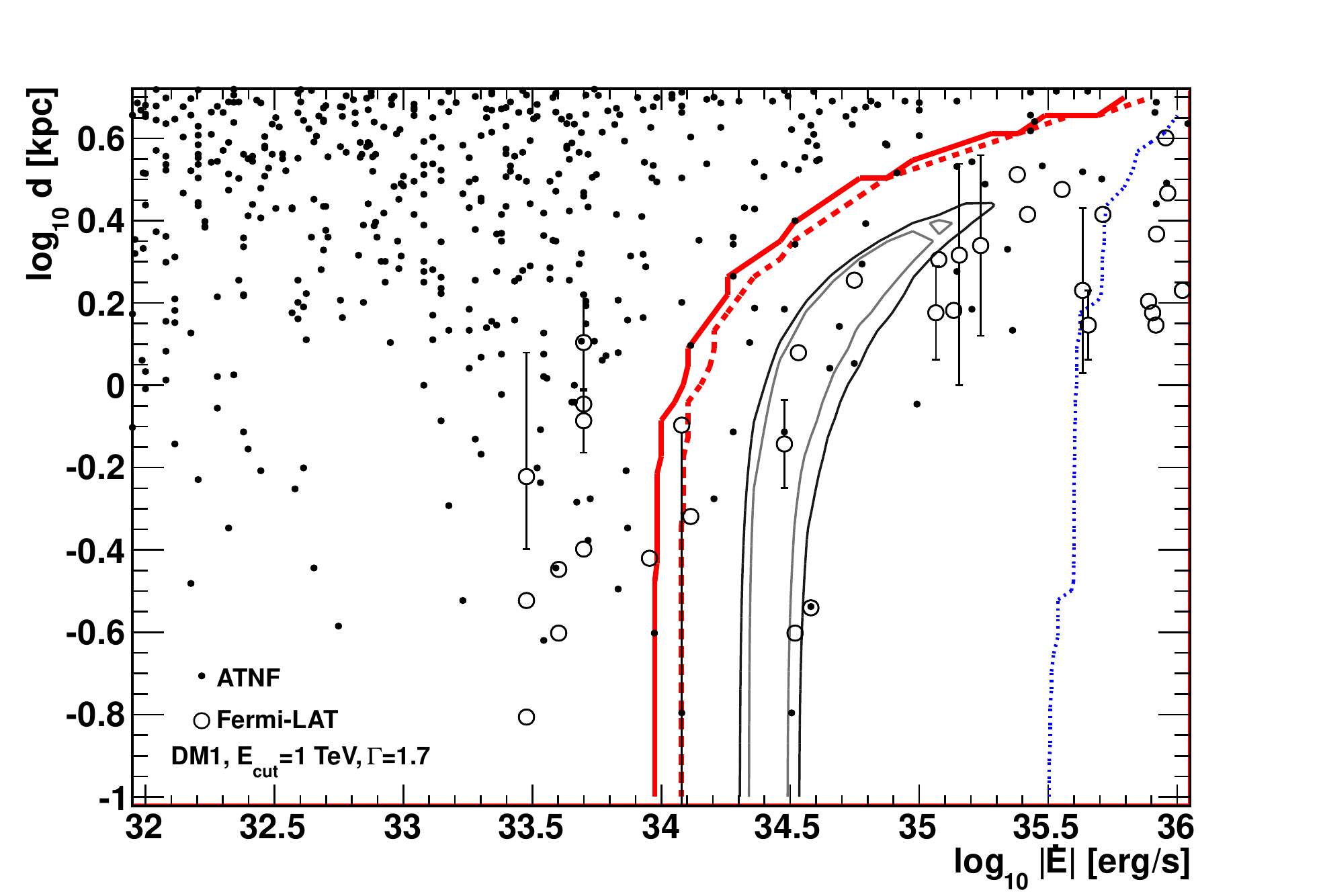}
 \includegraphics[width=9.cm,height=7.5cm]{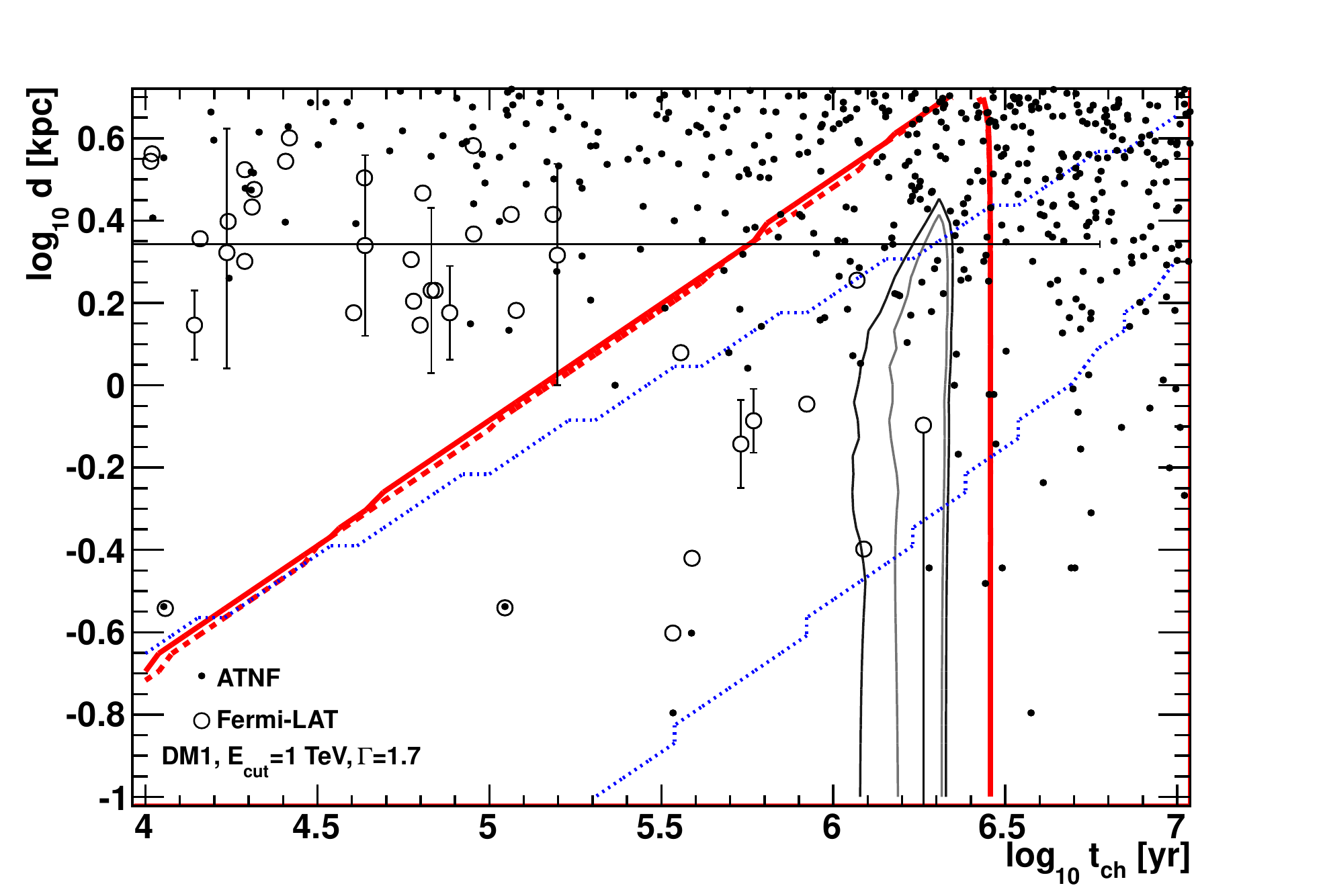}
 \caption{\fontsize{9}{9}\selectfont Regions in the pulsar parameter space compatible with the mock AMS-02  data shown in the left column of Figure \ref{figDMdata} corresponding to the fiducial model DM1. The thin inner (outer) solid contour corresponds to an $1-\alpha=50\%$ (99\%) confidence level goodness of fit. The thick solid (dashed) red lines individuate the regions where a single pulsar $-$ plus background $-$ contributes $>$50\% of the positron fraction (flux) at $\sim86$ GeV. The dotted blue line delimits the 2$\sigma$ $e^{\pm}$ anisotropy reach of Fermi-LAT after 5 years. Also shown are the ATNF catalogue \cite{atnf} (filled dots) and Fermi-LAT pulsars \cite{FermiCat,Fermi8} (open circles). Notice that not all pulsars visually within the thin contours are actually inside the corresponding three-dimensional confidence regions since there is a third, hidden dimension.}\label{figfit1}
\end{figure*}

\par For each DM fiducial model, ``mock'' AMS-02 measurements of the cosmic ray electronic component are generated according to the capabilities outlined in section \ref{secexp}. The resulting positron fraction and positron flux are shown in Figure \ref{figDMdata}. Using these data (we consider only energies above 10 GeV to minimise solar modulation dependence), the pulsar hypothesis is tested by performing a fixed grid scan within the ranges $|\dot{E}|=10^{32}-10^{36}$ erg/s, $d=0.1-5$ kpc and $t_{ch}=10^4-10^7$ yr. In this work we are interested in the goodness of fit and not in extracting best-fit parameters and the corresponding uncertainties. Therefore, we compute the $\chi^2$ for each pulsar parameter set and draw the contours outside which parameters are excluded at $1-\alpha=99\%$ or $50\%$ confidence level (using as degrees of freedom the number of available energy bins above 10 GeV). The two dimensional contours presented in the following are obtained by selecting the minimum values of $\chi^2$ along the hidden dimension. Figure \ref{figfit1} shows the results for DM1, $\Gamma=1.7$ and $E_{cut}=1$ TeV, as well as ATNF and Fermi-LAT pulsars. Also indicated are the regions where a single pulsar $-$ plus background $-$ contributes $>50\%$ to the positron fraction (thick solid red) and positron flux (thick dashed red) at $\sim$86 GeV, along with the pulsars producing anisotropies visible at 2$\sigma$ to Fermi-LAT after 5 years (dotted blue). The spectra produced by the best fit pulsar parameter set is shown by the dotted line in the left column of Figure \ref{figDMdata} -- the distinction between DM and pulsars seems virtually impossible with AMS-02 mock data. Several comments are in order here. First, the benchmark DM1 features a cutoff at 100 GeV which is well inside AMS-02 range and hence in this case a precise measurement of the spectral feature is anticipated. This is an optimistic scenario where the compatible regions in the pulsar parameter space are tight, as shown in Figure \ref{figfit1}. In particular, the contours restrict very effectively the values for the pulsar characteristic age $t_{ch}$ needed to mimic the DM signal. This is because $t_{ch}$ fixes the maximal energy $E_{max}\simeq 1/(b_0 t_{ch})$ that induces a rather sharp cutoff (in this case where $E_{cut}>E_{max}$). The normalisation is instead given by $|\dot{E}|$ and $d$: more distant pulsars require larger energy inputs to produce the same propagated spectrum. Such a behaviour $-$ illustrated in the top right plot of Figure \ref{figfit1} $-$ breaks when $d\ll r_{dif}$ (check equation \eqref{ne2}) in which case $|\dot{E}|$ alone fixes the normalisation.

\begin{figure*}
 \centering
\hspace{-1.0cm}
 \includegraphics[width=0.345\textwidth,height=5.5cm]{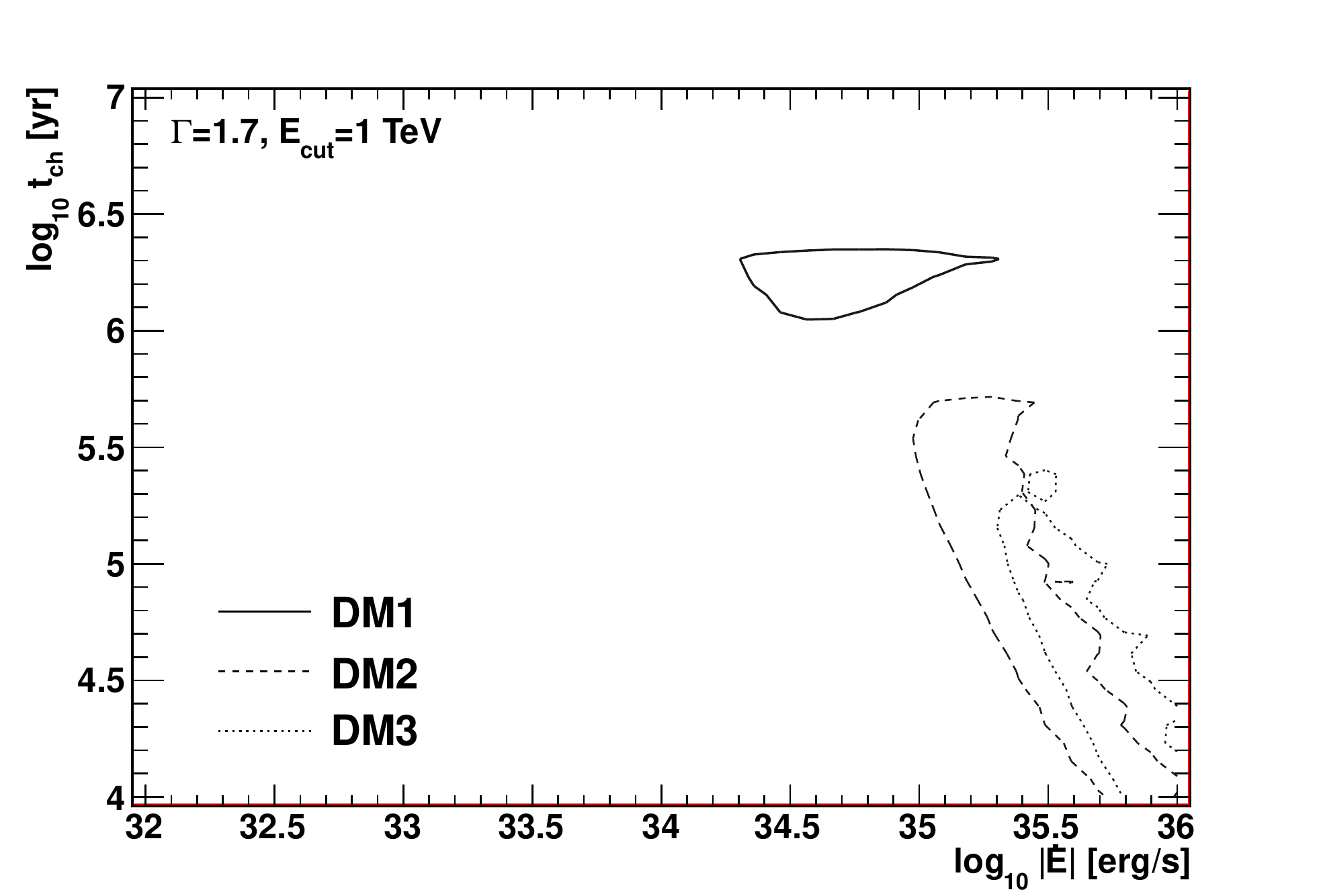}
 \includegraphics[width=0.345\textwidth,height=5.5cm]{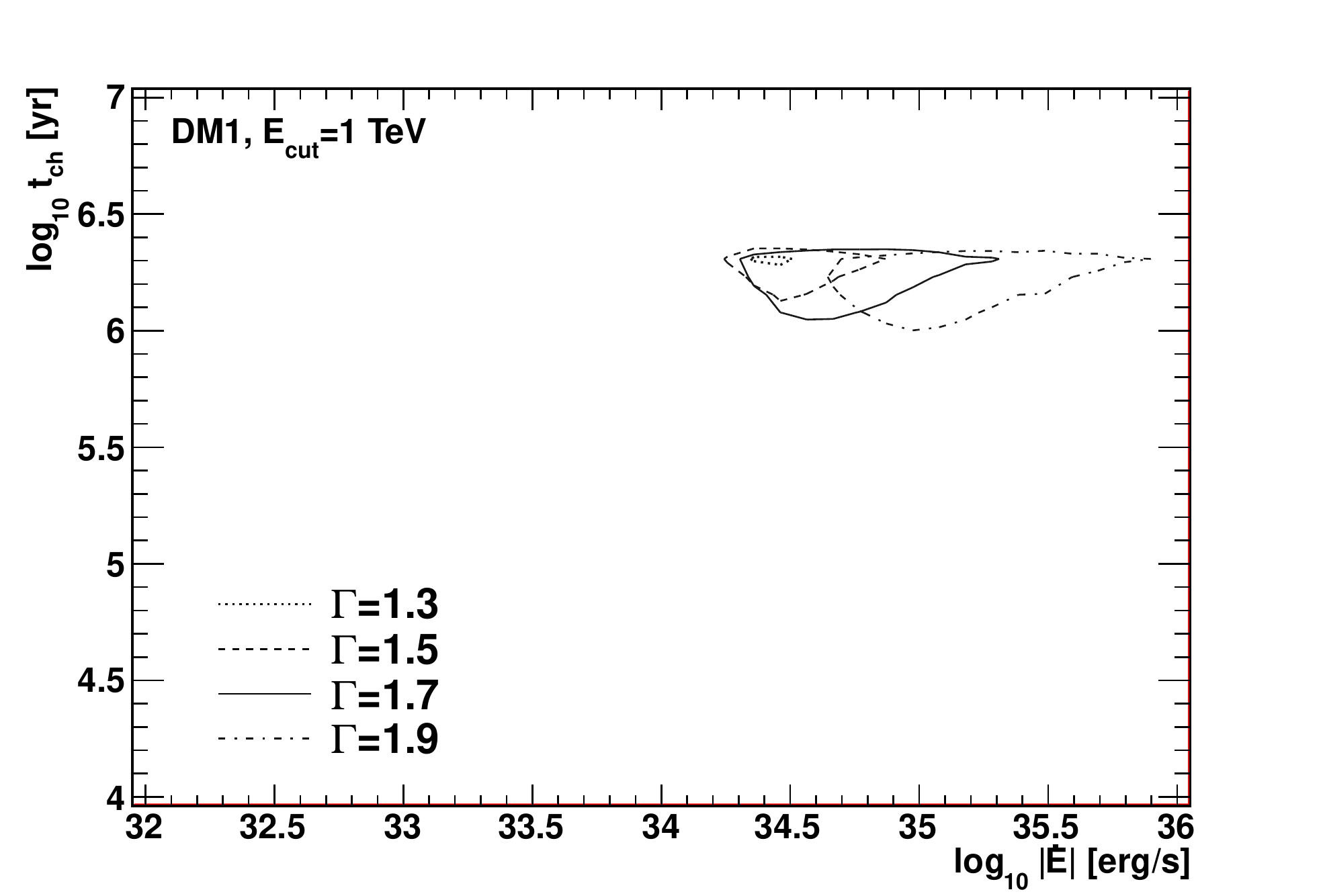}
 \includegraphics[width=0.345\textwidth,height=5.5cm]{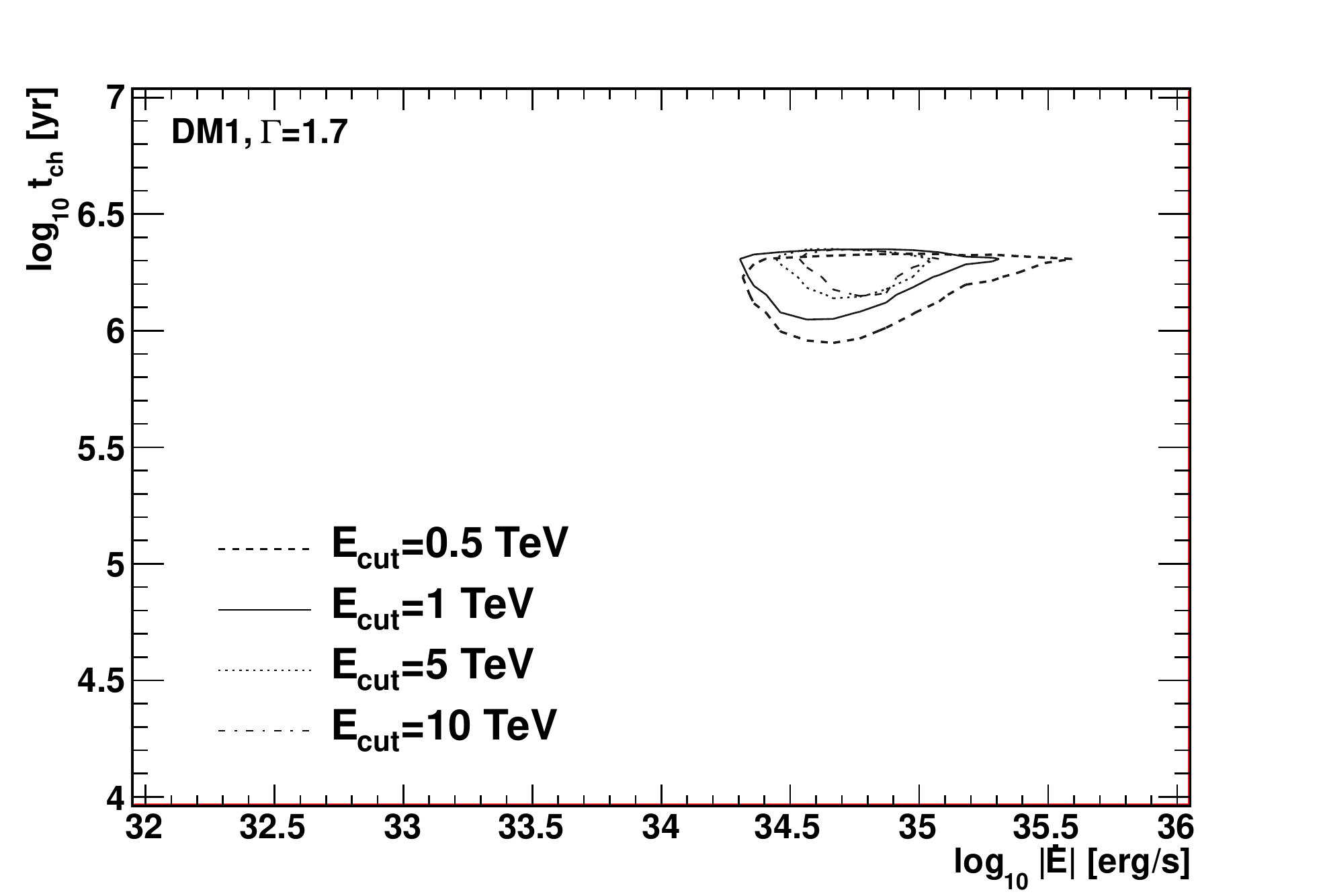}
 \caption{\fontsize{9}{9}\selectfont The 99\% CL region compatible with the mock AMS-02 data in Figure \ref{figDMdata} for different DM models (left), $\Gamma$ (central) and $E_{cut}$ (right).}\label{figdepend}
\end{figure*}

\par In Figure \ref{figdepend} we show how the confidence regions in the plane $t_{ch}$ vs.~$|\dot{E}|$ are affected by the choice of the spectral index $\Gamma$, the cutoff energy $E_{cut}$ and the DM fiducial model. As it is evident from the left plot of this figure, data featuring cutoffs at higher energy select younger pulsars (since $E_{max}\propto t_{ch}^{-1}$), and are compatible with larger portions of the pulsar parameter space because of the larger experimental uncertainties at high energies. On the other hand, changing the spectral index or the cutoff energy changes less importantly the allowed regions in the pulsar parameter space.

\begin{table*}
\centering
\fontsize{9}{9}\selectfont
\begin{tabular}{ccc|cccc}
\hline
\hline
benchmark & $\Gamma$ & $E_{cut}$ [GeV] & $N_1^{50}$ & $N_1^{99}$ & $N_2$ & $N_3$ \\
\hline
DM1 & 1.3 & 1000 & $\quad$ 0 $\quad$ & $\quad$  0 $\quad$ & $\quad$ 5 $\quad$ & $\quad$ 3 $\quad$ \\
DM1 & 1.5 & 1000 & 0 & 0 & 5 & 5 \\
DM1 & 1.7 & 1000 & 0 & 0 & 5 & 3 \\
DM1 & 1.9 & 1000 & 0 & 0 & 1 & 1 \\
\hline
DM1 & 1.7 & 500 & 0 & 1 &  4 & 4 \\
DM1 & 1.7 & 5000 & 0 & 0 & 1 & 1 \\
DM1 & 1.7 & $10^4$ & 0 & 0 & 1 & 1 \\
\hline
DM2 & 1.5 & 1000 & 0 & 0 & 8 & 6 \\
DM2 & 1.7 & 1000 & 0 & 0 & 6 & 5 \\
DM2 & 1.9 & 1000 & 0 & 0 & 3 & 1 \\
\hline
DM3 & 1.5 & 1000 & 0 & 0 & 8 & 6 \\
DM3 & 1.7 & 1000 & 0 & 0 & 8 & 5 \\
DM3 & 1.9 & 1000 & 0 & 0 & 4 & 2 \\
\hline
\end{tabular}
\caption{\fontsize{9}{9}\selectfont The number of known pulsars (from ANTF and Fermi-LAT catalogues) producing good fits to the mock AMS-02 data in Figure \ref{figDMdata} for different combinations of $\Gamma$ and $E_{cut}$. $N_1^{50,99}$ represent the number of catalogue pulsars found within the 50\%, 99\% CL contours, while $N_2$ ($N_3$) is the number of pulsars that contribute individually $-$ plus background $-$ more than $50\%$ of the positron fraction (flux) at $\sim$86 GeV.}\label{tab2}
\end{table*}

\par We summarise in Table \ref{tab2} the number of catalogue pulsars producing good fits to the mock data and the ones expected to contribute non-negligibly to the electron/positron anticipated spectrum. Notice that the ATNF and Fermi-LAT catalogues have common objects (though with different derived properties) and so the figures in Table \ref{tab2} report the number of different pulsars found in each case. The bottom line of this calculation is that, from the phenomenological viewpoint, it is possible to mimic a DM-like spectrum with single pulsars, even in the extreme case of direct DM annihilations into electron-positron pairs where a sharp cutoff is present. By construction, the mock data shown in Figure \ref{figDMdata} are  better fit by DM than pulsars, but, statistically speaking, one cannot exclude at a high confidence level the regions inside the contours drawn in Figures \ref{figfit1} and \ref{figdepend}. However, within these regions we find very few or none known catalogue pulsars as indicated in Table \ref{tab2}. In fact, only for the case DM1, $\Gamma=1.7$,  $E_{cut}=500\,\mathrm{GeV}$ does one find a pulsar setup that can reproduce the mock data in a satisfactory manner. This may be interpreted as an argument against the pulsar hypothesis, even though catalogues are likely incomplete since pulsars emit electromagnetic radiation in a directional fashion and thus only a fraction of these objects may be observed from Earth -- therefore the pulsar hypothesis cannot be ruled out on this basis. In order to estimate the incompleteness of catalogues, one may compute the so-called beaming fraction defined as the fraction of sky each pulsar beam sweeps, $F(\alpha\neq(0,\pi),\Delta\theta)=\frac{1}{4\pi}\int_{\Delta\Omega}{d\Omega}=\frac{1}{2}\left[cos\left(\alpha-\Delta\theta/2\right)-cos\left(\alpha+\Delta\theta/2\right)\right]$, where $\alpha$ is the angle between rotation and magnetic axes, $\Delta\theta/2$ is the half-width of the beam and we have considered top-hat beams (see e.g.~\cite{Ravi}). Assuming an isotropic distribution for $\alpha$, the mean beaming fraction reads $\langle F\rangle(\Delta\theta)=\int_{0}^{\pi/2}{d\alpha \, F(\alpha,\Delta\theta)sin\alpha}=\frac{\pi}{4}sin\left(\frac{\Delta\theta}{2}\right)$. The width of the beam $\Delta\theta$ is, of course, dependent on the emission mechanism. Radio beams, for instance have typical half-openings $\Delta\theta/2=5.8^{\circ}(P/\textrm{s})^{-1/2}$ \cite{Rankin} which for a fiducial period $P\sim 0.1$ s gives $\Delta\theta/2\simeq 18^{\circ}$ and $\langle F_r\rangle\sim 0.24$. However, values $F_r\sim 1$ have been found by studying different populations of detected $\gamma$-ray and radio pulsars \cite{Ravi} indicating wide radio emission. As for $\gamma$-rays, beaming fractions in the literature vary in the range $F_g\sim0.1-1$ \cite{Watters}. Now, if the distribution of nearby pulsars is assumed isotropic, the ratio of total to observed numbers of objects should be given by $1/F$, or $\sim$1--4 (1--10) using the above-stated radio ($\gamma$-ray) beaming fractions. These figures do not include distance selection effects, but recall that in the present work we are interested in nearby, high-luminosity objects.

\begin{figure*}[ht]
 \centering
\hspace{-1cm}
 \includegraphics[width=7.5cm,height=7.5cm]{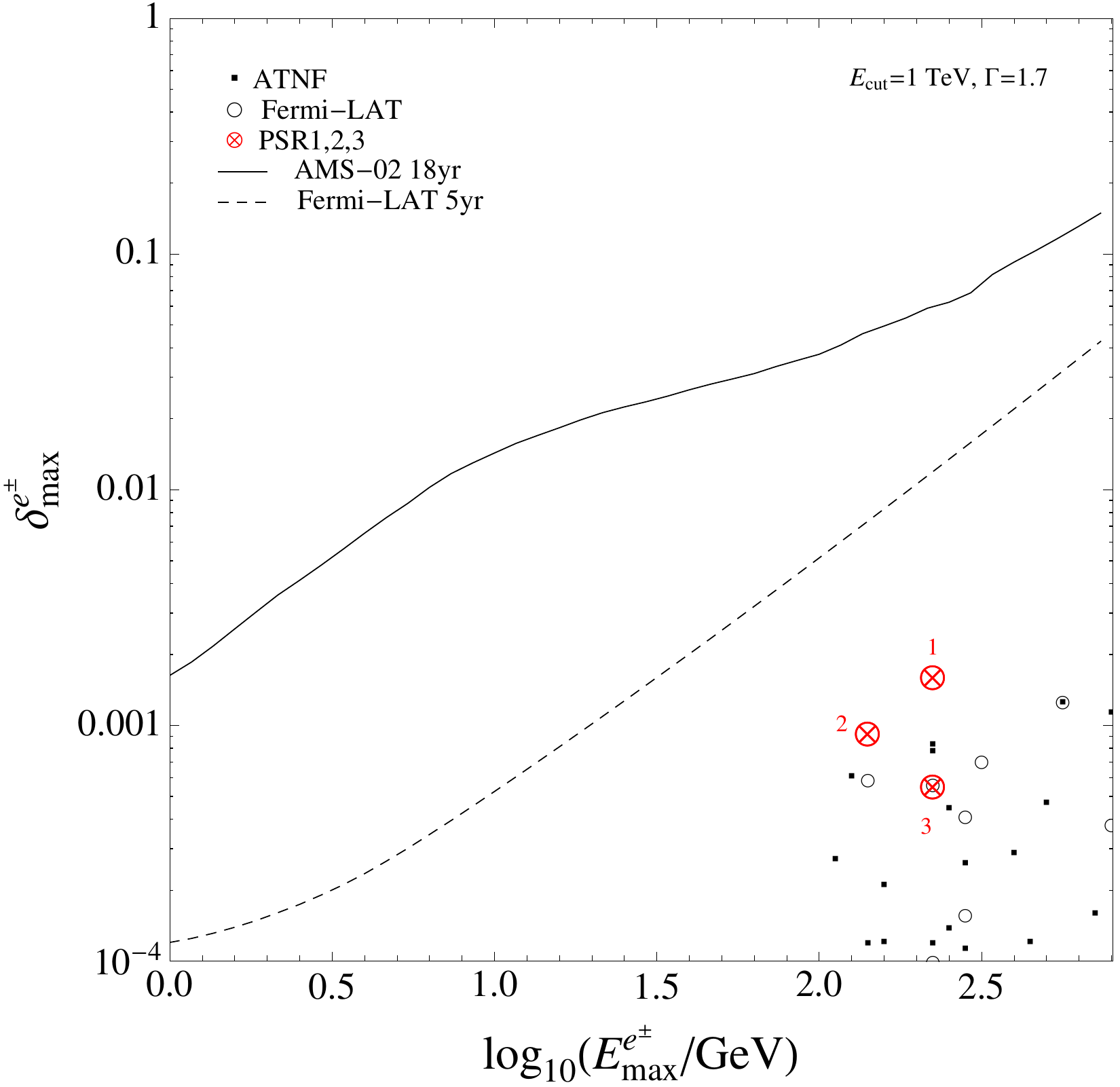}
 \includegraphics[width=7.5cm,height=7.5cm]{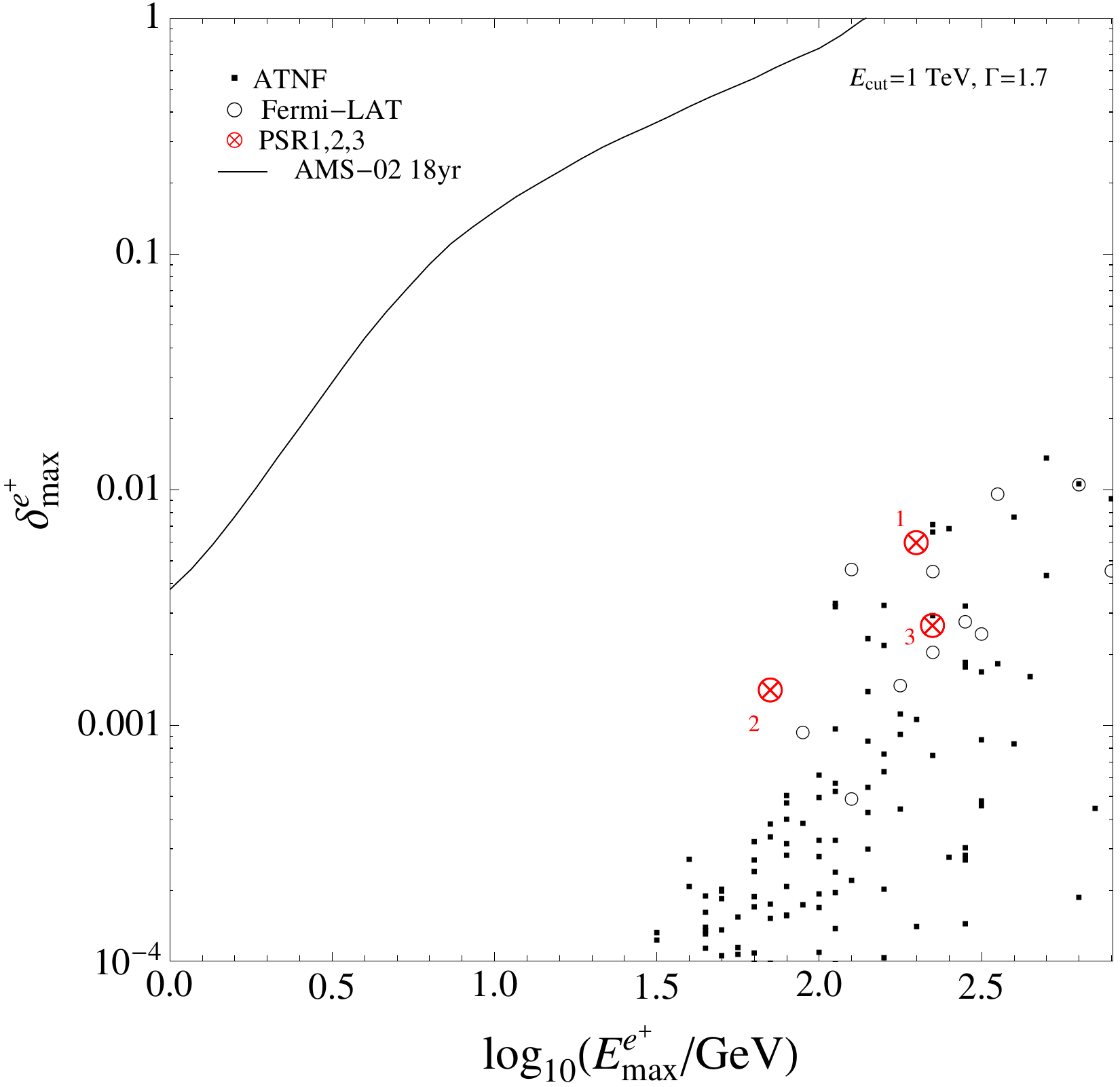}
 \caption{\fontsize{9}{9}\selectfont The maximum anisotropy produced by ATNF (filled dots) and Fermi-LAT (open circles) pulsars with $t_{ch}=10^4-10^7$ yr as a function of the energy at which such anisotropy is attained. Here, the values $\Gamma=1.7$, $E_{cut}=1000$ GeV and $f=0.9$ (besides $\tau_0=10^4$ yr and $\eta_{e^{\pm}}=0.4$) were assumed for all catalogue pulsars. The fiducial setups in Table \ref{tab3} are shown by red circled crosses. In the left (right) frame we show the $e^{\pm}$ ($e^+$) maximum anisotropy. The AMS-02 18 years (Fermi-LAT 5 years) 2$\sigma$ anisotropy reach is plotted in solid (dashed). Notice that $E_{max}^{e^{\pm}}$ and $E_{max}^{e^+}$ are different from the maximal energy allowed by losses, $E_{max}$.}\label{figanis}
\end{figure*}

\par Another difficulty with the single pulsar explanation evident from Table \ref{tab2} arises from the fact that at least a few known pulsars should contribute (together with the background) more that 50\% of the positron fraction and positron flux at a given energy ($\sim$ 86 GeV). Therefore, in order to accommodate the cutoffs of Figure \ref{figDMdata} with a single pulsar, one also needs to explain why a few known pulsars produce less electron-positron pairs than expected. A caveat to this argument is the fact that a rather high efficiency $\eta_{e^{\pm}}=0.4$ is being used. For the reference case DM1, $\Gamma=1.7$, $E_{cut}=1$ TeV, we redid the calculations with $\eta_{e^{\pm}}=0.04$ and found no known pulsar contributing significantly to the $e^{+}$ flux or fraction, which indeed weakens the above-mentioned reasoning.

\par It has been proposed in the literature that the pulsar origin of the cosmic ray lepton excess could be tested by means of anisotropy measurements. Unfortunately, however, we find no catalogue pulsars producing $e^{\pm}$ or $e^{+}$ anisotropy levels visible to Fermi-LAT in 5 years or AMS-02 in 18 years. This conclusion is clear in Figure \ref{figanis} where the experimental reaches have been plotted along with the maximal anisotropy $-$ attained at energy $E_{max}^{e^{\pm}}$ or $E_{max}^{e^{+}}$ $-$ featured by known pulsars with ages $t_{ch}=10^4-10^7$ yr. Obviously it is possible that a low-energy anisotropy is visible while the maximal value goes undetected, but we do not consider such case since it would not be a strong evidence for singling out a particular pulsar. In this figure the references values $\Gamma=1.7$ and $E_{cut}=1000$ GeV were assumed, but we have also checked that the expected anisotropies from known pulsars are below the experimental reaches for all the combinations of parameters $(\Gamma,E_{cut})$ listed in Table \ref{tab2}. Notice that the recent 3$\sigma$ $e^{\pm}$ dipole anisotropy upper limits from Fermi-LAT \cite{fermianis} range from $\sim 10^{-2}$ at 100 GeV to $\sim 10^{-1}$ at 500 GeV $-$ somewhat in between the dashed and solid curves in the left frame of Figure \ref{figanis}. Note as well that the anisotropies we find are significantly smaller than in other works (see e.g.~\cite{Grasso2010}) because we are using an $e^{\pm}$ ``background'' tuned to roughly match Fermi-LAT data and we are not letting the catalogue pulsars output $|\dot{E}|$ vary to explain the lepton excess.

\begin{figure*}[htp]
 \centering
 \includegraphics[width=7.5cm,height=7.5cm]{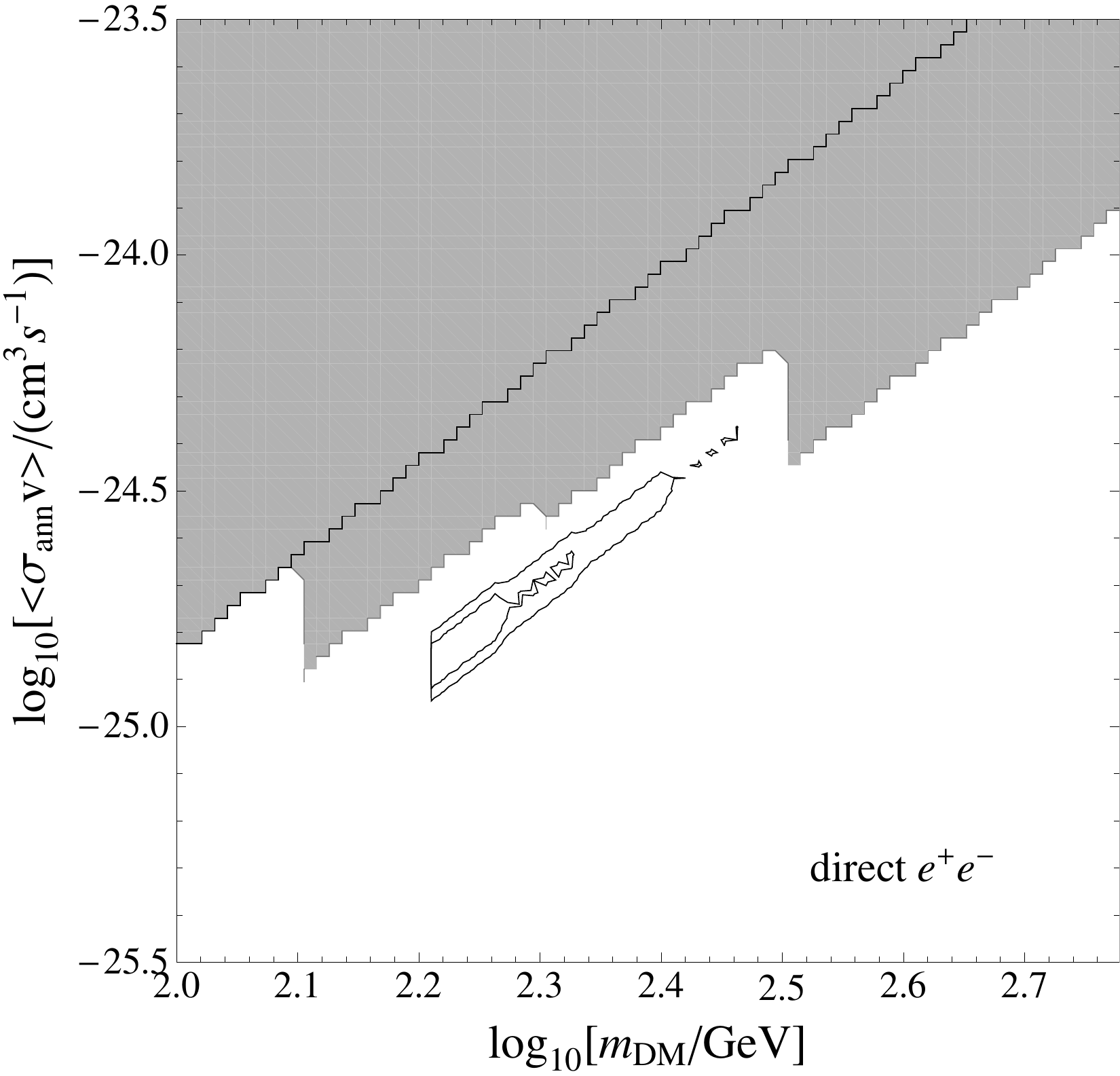}
 \includegraphics[width=7.5cm,height=7.5cm]{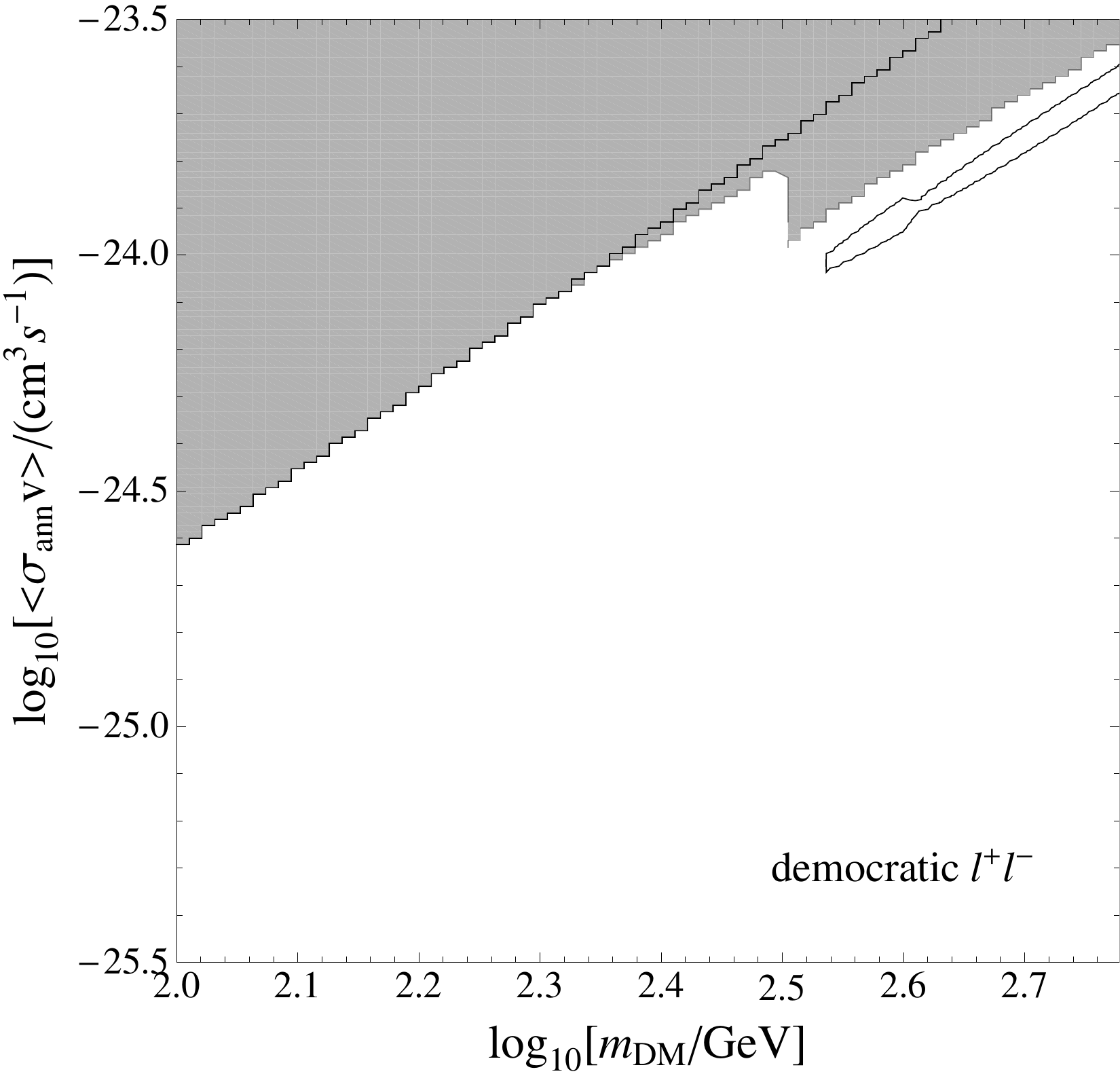}
 \caption{\fontsize{9}{9}\selectfont The regions in the dark matter parameter space compatible with the mock AMS-02 data assuming the three pulsar benchmarks of Table \ref{tab3}. The inner (outer) contour corresponds to an $1-\alpha=50\%$ (99\%) confidence level goodness of fit. In the left frame, direct annihilations into $e^+e^-$ are assumed and the solid contours correspond to PSR2; in this case no compatibility is found for PSR1 nor PSR3. In the right frame, democratic annihilations into leptons are assumed and the solid contour ($1-\alpha=99\%$) corresponds to PSR1; in this case no compatibility is found for PSR2 nor PSR3. In both frames, the shaded region is excluded at 3$\sigma$ by PAMELA positron fraction or Fermi-LAT/H.E.S.S.~electron plus positron flux, being that the portion above the upper solid line is excluded by PAMELA.}\label{figfit2}
\end{figure*}

\vspace{0.5cm}
\par We now turn to the inverse problem: we assume a single pulsar as the source of high-energy electrons and positrons, and evaluate the prospects for distinguishing this scenario from that of DM annihilations. The three fiducial sets of pulsar properties shown in Table \ref{tab3} and Figure \ref{figposfracPSR} were adopted, each featuring a different cutoff sharpness. Applying a procedure very similar to the previous case, we generate mock data for the positron fraction and positron flux in the pulsar scenario and test the DM hypothesis. Direct annihilations into $e^{\pm}$ and democratic annihilations into leptons are both considered, and $m_{DM}$ and $\langle \sigma_{ann} v \rangle$ are treated as free parameters. The resulting $1-\alpha=99\%,50\%$ contours are presented in Figure \ref{figfit2}. The shaded regions are excluded at 3$\sigma$ by present data on the positron fraction (PAMELA) and/or the electron plus positron flux (Fermi-LAT, H.E.S.S.). As evident from Figure \ref{figfit2}, DM models with direct annihilations into electron-positron pairs can mimic the spectrum produced by the benchmark PSR2, but not the first and third cases that present milder cutoffs. Analogously, democratic DM annihilations into charged leptons cannot be ruled out as explanation of the spectrum induced by PSR1, but do not explain a sharp cutoff as the one produced by PSR2. For the sake of clarity we have not considered further annihilation channels that may ease the fit to mild pulsar-like cutoffs. In any case, for the representative pulsar cases, one can always have well-fit DM spectra either with annihilations into $e^{\pm}$ or democratic leptons. Of course, as extensively discussed in the literature and already pointed out in Section \ref{secintro}, the large cross-sections highlighted in Figure \ref{figfit2} are excluded by or in tension with other data.

\vspace{0.5cm}
\par Throughout our work we have assumed (almost) ``perfect data'', i.e.~mock data presenting no fluctuations with respect to the ``true'' observables (but including the smearing due to the energy resolution of the instrument). This is of course a quite optimistic assumption regarding the ability to discriminate different sources and therefore strengthens our conclusions. Nevertheless, for the benchmark DM1 we have generated mock data by drawing the flux in each energy bin from a Gaussian distribution of mean given by the flux corresponding to $N_X$ in equation \eqref{NX} and standard deviation as explained in Section \ref{secexp}. In that case, for a single realisation, the true model yields a good (but not perfect) fit to the mock data with $\chi_{DM}^2/n_{dof}=40.7/58$ ($p=95.6\%$), as opposed to $\chi_{DM}\sim0$ using ``perfect data''. On the other hand, fixing $\Gamma=1.7$ and $E_{cut}=1$ TeV, we find that the best pulsar parameter set provides a chi-square $\chi_{pul}^2/n_{dof}=42.1/58$ ($p=94.2\%$), which cannot be ruled out with any reasonable statistical significance. We have also checked that the 99\% CL contour in the pulsar parameter space obtained using mock data with fluctuations encompasses a region similar (but smaller) than the one shown in Figures \ref{figfit1} and \ref{figdepend}. In the future, a dedicated study of the performance of AMS-02 in detecting electrons and positrons may allow a more realistic analysis, but we stress here that such would reinforce the idea that rejecting the pulsar hypothesis in case of a DM-like spectrum (or vice-versa) will be highly challenging.

\section{Conclusions}\label{secconc}

\par In this work we have studied the capability of future precision $e^{\pm}$ measurements to constrain possible origins of the cosmic-ray lepton excess, focusing on the dark matter and pulsar interpretations. In order to model upcoming experimental capabilities, the performance of AMS-02 was estimated in a realistic way, including both systematic and statistical uncertainties. We have considered the case in which a cutoff in the electron/positron spectrum, produced either by DM annihilations or by acceleration in a pulsar, is observed, and quantified the ability of AMS-02 to reject the wrong hypothesis. In particular, the first scenario studied was the case where the excess is due to DM direct annihilations to $e^{\pm}$ pairs, for three different values of the DM mass ($m_{DM}=100,\,300,\,500\,\mathrm{GeV}$). Even if the DM-induced spectral cutoff is quite sharp, it turns out that it is always possible to find values for the pulsar spin-down luminosity, distance and characteristic age that produce an electron-positron spectrum that would be compatible, within the experimental uncertainties, with the DM one. However, such would require quite specific values of the pulsar luminosity and age. In fact, in nearly all cases, none of the kwown pulsars within the ATNF catalogue, nor of the Fermi-LAT pulsars, satisfies these requirements, although it should be taken into account that catalogues are likely incomplete. Then, we have considered the symmetric case where the excess is produced by a pulsar, and tried to fit the anticipated data with DM directly annihilating either to $e^+/e^-$ pairs or democratically (i.e. 33\% of the time each) to $e^+/e^-$, $\mu^+/\mu^-$ and $\tau^+/\tau^-$. Within the framework of each model, DM was treated in an as much model-independent way as possible, taking the DM mass and annihilation cross section as independent parameters; nevertheless, the ``democratic'' case can be seen as somewhat representative of KK and leptophilic DM models. We find that the possibility to reproduce the pulsar-induced spectrum, as well as the volume in the DM parameter space (in other words the amount of fine tuning) required to do so critically depends on the shape of the cutoff, and thus on the characteristics of the pulsar originating the excess. In any case, generically speaking, it is possible to mimic a pulsar-like spectrum with DM annihilations, even though the required cross-sections are usually in tension with other data as extensively discussed in the literature.

\par Let us point out that our results were obtained in favourable conditions regarding a possible source discrimination, namely by fixing the propagation setup. Including propagation uncertainties would likely worsen the prospects for discrimination. The main conclusion of our work is therefore that future $e^{\pm}$ data will likely be insufficient to discriminate between the dark matter and the single pulsar interpretations of the cosmic-ray lepton excess. One caveat to this statement would be the detection of several bumps in the electron-positron spectrum at high energies that could be associated to the contribution of several nearby pulsars, and that would be difficult to mimic with dark matter annihilations or decays (see e.g.~\cite{cholis}). In the future, complementary data, or a better understanding of both sources and cosmic-ray propagation, may contribute to a better discrimination power than presented here.

\vspace{2cm}

{\it Acknowledgements:} We wish to thank Pasquale Blasi, Stefano Profumo and Antje Putze for useful suggestions. MP is supported by Funda\c{c}\~{a}o para a Ci\^encia e Tecnologia (Minist\'erio da Ci\^encia, Tecnologia e Ensino Superior).


\begin{thebibliography}{99}


\bibitem{Casadei}
  D.~Casadei and V.~Bindi,
  Astrophys.\ J.\  {\bf 612} (2004) 262.

\bibitem{Longair2}
  M.~S.~Longair,
  ``High-energy astrophysics. Vol. 2: Stars, the galaxy and the interstellar
  medium,''
{\it  Cambridge University Press; 2nd edition (1994).}

\bibitem{caprice94}
  M.~Boezio {\it et al.}
  Astrophys.\ J.\  {\bf 532} (2000) 653.

\bibitem{heat9495}
  S.~W.~Barwick {\it et al.}  [HEAT Collaboration],
  Astrophys.\ J.\  {\bf 482} (1997) L191
  [arXiv:astro-ph/9703192].


\bibitem{heat00}
  J.~J.~Beatty {\it et al.},
  Phys.\ Rev.\ Lett.\  {\bf 93} (2004) 241102
  [arXiv:astro-ph/0412230].



\bibitem{ATIC}
  J.~Chang {\it et al.},
  Nature {\bf 456} (2008) 362.

\bibitem{PPBBETS}
  S.~Torii {\it et al.}  [PPB-BETS Collaboration],
  arXiv:0809.0760 [astro-ph].




\bibitem{PAMELA}
  O.~Adriani {\it et al.}  [PAMELA Collaboration],
  Nature {\bf 458} (2009) 607
  [arXiv:0810.4995 [astro-ph]].

\bibitem{PAMELA10}
  O.~Adriani {\it et al.},
  Astropart.\ Phys.\  {\bf 34} (2010) 1
  [arXiv:1001.3522 [Unknown]].



\bibitem{fermi}
  A.~A.~Abdo {\it et al.}  [The Fermi LAT Collaboration],
  Phys.\ Rev.\ Lett.\  {\bf 102} (2009) 181101
  [arXiv:0905.0025 [astro-ph.HE]].

\bibitem{fermilat10}
  M.~Ackermann {\it et al.}  [Fermi LAT Collaboration],
  arXiv:1008.3999 [astro-ph.HE].



\bibitem{hess08}
  F.~Aharonian {\it et al.}  [H.E.S.S. Collaboration],
  Phys.\ Rev.\ Lett.\  {\bf 101} (2008) 261104
  [arXiv:0811.3894 [astro-ph]].

\bibitem{hess09}
  F.~Aharonian {\it et al.}  [H.E.S.S. Collaboration],
  Astron.\ Astrophys.\  {\bf 508} (2009) 561
  [arXiv:0905.0105 [astro-ph.HE]].

\bibitem{Delahaye08}
  T.~Delahaye, F.~Donato, N.~Fornengo, J.~Lavalle, R.~Lineros, P.~Salati and R.~Taillet,
  Astron.\ Astrophys.\  {\bf 501} (2009) 821
  [arXiv:0809.5268 [astro-ph]].

\bibitem{Serpico}
  P.~D.~Serpico,
  Phys.\ Rev.\  D {\bf 79} (2009) 021302
  [arXiv:0810.4846 [hep-ph]].


\bibitem{Atoyan}
 A.~M.~Atoyan, F.~A.~Aharonian and H.~J.~V\"olk  ,
 Phys.\ Rev.\ D 52, 3265–3275 (1995).


\bibitem{cirelli}
  M.~Cirelli, M.~Kadastik, M.~Raidal and A.~Strumia,
  Nucl.\ Phys.\  B {\bf 813} (2009) 1
  [arXiv:0809.2409 [hep-ph]].

\bibitem{Tran}
  A.~Ibarra and D.~Tran,
  JCAP {\bf 0902} (2009) 021
  [arXiv:0811.1555 [hep-ph]].
  
\bibitem{Hisano:2004ds}
  J.~Hisano, S.~Matsumoto, M.~M.~Nojiri and O.~Saito,
  Phys.\ Rev.\  D {\bf 71}, 063528 (2005)
  [arXiv:hep-ph/0412403].
  
\bibitem{Cirelli:2007xd}
  M.~Cirelli, A.~Strumia and M.~Tamburini,
  Nucl.\ Phys.\  B {\bf 787}, 152 (2007)
  [arXiv:0706.4071 [hep-ph]].

\bibitem{ArkaniHamed:2008qn}
  N.~Arkani-Hamed, D.~P.~Finkbeiner, T.~R.~Slatyer and N.~Weiner,
  Phys.\ Rev.\  D {\bf 79}, 015014 (2009)
  [arXiv:0810.0713 [hep-ph]].

\bibitem{Pospelov:2008jd}
  M.~Pospelov and A.~Ritz,
  Phys.\ Lett.\  B {\bf 671}, 391 (2009)
  [arXiv:0810.1502 [hep-ph]].
    
\bibitem{Lattanzi:2008qa}
  M.~Lattanzi and J.~I.~Silk,
  Phys.\ Rev.\  D {\bf 79}, 083523 (2009)
  [arXiv:0812.0360 [astro-ph]].
  
\bibitem{MarchRussell:2008tu}
  J.~D.~March-Russell and S.~M.~West,
  Phys.\ Lett.\  B {\bf 676}, 133 (2009)
  [arXiv:0812.0559 [astro-ph]].

\bibitem{Taoso}
  G.~Bertone, M.~Cirelli, A.~Strumia and M.~Taoso,
  JCAP {\bf 0903} (2009) 009
  [arXiv:0811.3744 [astro-ph]].
 
\bibitem{Profumo:2009uf}
  S.~Profumo and T.~E.~Jeltema,
  JCAP {\bf 0907}, 020 (2009)
  [arXiv:0906.0001 [astro-ph.CO]].
    
\bibitem{Belikov:2009cx}
  A.~V.~Belikov and D.~Hooper,
  Phys.\ Rev.\  D {\bf 81}, 043505 (2010)
  [arXiv:0906.2251 [astro-ph.CO]].

\bibitem{galli}
  S.~Galli, F.~Iocco, G.~Bertone and A.~Melchiorri,
  Phys.\ Rev.\  D {\bf 80} (2009) 023505
  [arXiv:0905.0003 [astro-ph.CO]].

\bibitem{fink}
  T.~R.~Slatyer, N.~Padmanabhan and D.~P.~Finkbeiner,
  Phys.\ Rev.\  D {\bf 80} (2009) 043526
  [arXiv:0906.1197 [astro-ph.CO]].


\bibitem{Iocco}
  M.~Cirelli, F.~Iocco and P.~Panci,
  JCAP {\bf 0910} (2009) 009
  [arXiv:0907.0719 [astro-ph.CO]].

\bibitem{Delahaye2}
  F.~Donato, D.~Maurin, P.~Brun, T.~Delahaye and P.~Salati,
  Phys.\ Rev.\ Lett.\  {\bf 102} (2009) 071301
  [arXiv:0810.5292 [astro-ph]].

\bibitem{Buesching}
  I.~Buesching, O.~C.~de Jager, M.~S.~Potgieter and C.~Venter,
  arXiv:0804.0220 [astro-ph].

\bibitem{Hooper}
  D.~Hooper, P.~Blasi and P.~D.~Serpico,
  JCAP {\bf 0901} (2009) 025
  [arXiv:0810.1527 [astro-ph]].


\bibitem{Profumo}
  S.~Profumo,
  arXiv:0812.4457 [astro-ph].

\bibitem{Delahaye3}
  T.~Delahaye, J.~Lavalle, R.~Lineros, F.~Donato and N.~Fornengo,
  arXiv:1002.1910 [Unknown].

\bibitem{Gendelev}
  L.~Gendelev, S.~Profumo and M.~Dormody,
  JCAP {\bf 1002} (2010) 016
  [arXiv:1001.4540 [Unknown]].

\bibitem{Latronico}
  T.~Kamae {\it et al.},
  arXiv:1010.3477 [astro-ph.HE].



\bibitem{Blasi}
  P.~Blasi,
  Phys.\ Rev.\ Lett.\  {\bf 103} (2009) 051104
  [arXiv:0903.2794 [astro-ph.HE]].


\bibitem{BlasiSerpico}
  P.~Blasi and P.~D.~Serpico,
  Phys.\ Rev.\ Lett.\  {\bf 103} (2009) 081103
  [arXiv:0904.0871 [astro-ph.HE]].

\bibitem{Sarkar}
  P.~Mertsch and S.~Sarkar,
  Phys.\ Rev.\ Lett.\  {\bf 103} (2009) 081104
  [arXiv:0905.3152 [astro-ph.HE]].


\bibitem{HallHooper}
  J.~Hall and D.~Hooper,
  Phys.\ Lett.\  B {\bf 681} (2009) 220
  [arXiv:0811.3362 [astro-ph]].



\bibitem{atnf}
Manchester, R. N., Hobbs, G. B., Teoh, A. \& Hobbs, M., AJ, 129, 1993-2006 (2005).
 \href{}{www.atnf.csiro.au/research/pulsar/psrcat/}

\bibitem{FermiCat}
  A.~A.~Abdo {\it et al.}  [Fermi LAT collaboration],
  Astrophys.\ J.\ Suppl.\  {\bf 187} (2010) 460
  [arXiv:0910.1608 [Unknown]].


\bibitem{Fermi8}
  P.~M.~S.~Parkinson {\it et al.},
  arXiv:1006.2134 [Unknown].

\bibitem{ams02}
 \href{}{http://ams.cern.ch/}
 \href{}{http://www.ams02.org/}


\bibitem{heat9495e}
  M.~A.~DuVernois {\it et al.},
  Astrophys.\ J.\  {\bf 559} (2001) 296.

\bibitem{ams01}
  M.~Aguilar {\it et al.}  [AMS Collaboration],
  Phys.\ Rept.\  {\bf 366} (2002) 331
  [Erratum-ibid.\  {\bf 380} (2003) 97].

\bibitem{bets01}
  S.~Torii {\it et al.},
  Astrophys.\ J.\  {\bf 559} (2001) 973.

\bibitem{ams01efrac}
  M.~Aguilar {\it et al.}  [AMS-01 Collaboration],
  Phys.\ Lett.\  B {\bf 646} (2007) 145
  [arXiv:astro-ph/0703154].






\bibitem{grasso}
  D.~Grasso {\it et al.}  [FERMI-LAT Collaboration],
  Astropart.\ Phys.\  {\bf 32} (2009) 140
  [arXiv:0905.0636 [astro-ph.HE]].


\bibitem{galpropsite}
  \href{}{http://galprop.stanford.edu/web_galprop/galprop_home.html}


\bibitem{SM98}
  A.~W.~Strong and I.~V.~Moskalenko,
  Astrophys.\ J.\  {\bf 509} (1998) 212
  [arXiv:astro-ph/9807150].

\bibitem{GA68}
  L.~J.~Gleeson and W.~I.~Axford,
  Astrophys.\ J.\  {\bf 154} (1968) 1011.

\bibitem{creamhard}
  H.~S.~Ahn {\it et al.},
  Astrophys.\ J.\  {\bf 714} (2010) L89
  [arXiv:1004.1123 [astro-ph.HE]].

\bibitem{atichard}
  A.~D.~Panov {\it et al.},
  arXiv:astro-ph/0612377.

\bibitem{Lavallehard}
  J.~Lavalle,
  arXiv:1011.3063 [astro-ph.HE].

\bibitem{Putzehard}
  A.~Putze, D.~Maurin and F.~Donato,
  arXiv:1011.0989 [astro-ph.GA].

\bibitem{DonatoSerpicohard}
  F.~Donato and P.~D.~Serpico,
  arXiv:1010.5679 [astro-ph.HE].






\bibitem{VL2}
  J.~Diemand, M.~Kuhlen, P.~Madau, M.~Zemp, B.~Moore, D.~Potter and J.~Stadel,
  arXiv:0805.1244 [astro-ph].


\bibitem{multi}
  M.~Pato, L.~Pieri and G.~Bertone,
  Phys.\ Rev.\  D {\bf 80}, 103510 (2009)
  [arXiv:0905.0372 [astro-ph.HE]].


\bibitem{cosmo}
  R.~Catena, N.~Fornengo, M.~Pato, L.~Pieri and A.~Masiero,
  Phys.\ Rev.\  D {\bf 81}, 123522 (2010)
  [arXiv:0912.4421 [Unknown]].


\bibitem{Green}
  A.~M.~Green, S.~Hofmann and D.~J.~Schwarz,
  JCAP {\bf 0508}, 003 (2005)
  [arXiv:astro-ph/0503387].

\bibitem{Bringmann}
  T.~Bringmann,
  New J.\ Phys.\  {\bf 11} (2009) 105027
  [arXiv:0903.0189 [astro-ph.CO]].





\bibitem{Lavalle}
  J.~Lavalle, Q.~Yuan, D.~Maurin and X.~J.~Bi,
  arXiv:0709.3634 [astro-ph].


\bibitem{Delahaye4}
  T.~Delahaye, R.~Lineros, F.~Donato, N.~Fornengo and P.~Salati,
  Phys.\ Rev.\  D {\bf 77} (2008) 063527
  [arXiv:0712.2312 [astro-ph]].

  
  
\bibitem{Sjostrand:2000wi}
  T.~Sjostrand, P.~Eden, C.~Friberg, L.~Lonnblad, G.~Miu, S.~Mrenna and E.~Norrbin,
  Comput.\ Phys.\ Commun.\  {\bf 135}, 238 (2001)
  [arXiv:hep-ph/0010017].

\bibitem{Sjostrand:2006za}
  T.~Sjostrand, S.~Mrenna and P.~Z.~Skands,
  JHEP {\bf 0605}, 026 (2006)
  [arXiv:hep-ph/0603175].
  
\bibitem{Sjostrand:2007gs}
  T.~Sjostrand, S.~Mrenna and P.~Z.~Skands,
  Comput.\ Phys.\ Commun.\  {\bf 178}, 852 (2008)
  [arXiv:0710.3820 [hep-ph]].
  
\bibitem{Fox:2008kb}
  P.~J.~Fox and E.~Poppitz,
  Phys.\ Rev.\  D {\bf 79}, 083528 (2009)
  [arXiv:0811.0399 [hep-ph]].
  
\bibitem{Chun:2009zx}
  E.~J.~Chun, J.~C.~Park and S.~Scopel,
  JCAP {\bf 1002}, 015 (2010)
  [arXiv:0911.5273 [hep-ph]].
  
\bibitem{Hooper:2009fj}
  D.~Hooper and K.~M.~Zurek,
  Phys.\ Rev.\  D {\bf 79}, 103529 (2009)
  [arXiv:0902.0593 [hep-ph]].
  
\bibitem{Cholis:2008vb}
  I.~Cholis, L.~Goodenough and N.~Weiner,
  Phys.\ Rev.\  D {\bf 79}, 123505 (2009)
  [arXiv:0802.2922 [astro-ph]].
  
\bibitem{Cholis:2008qq}
  I.~Cholis, D.~P.~Finkbeiner, L.~Goodenough and N.~Weiner,
  JCAP {\bf 0912}, 007 (2009)
  [arXiv:0810.5344 [astro-ph]].
  

\bibitem{BlasiAmato}
  P.~Blasi and E.~Amato,
  arXiv:1007.4745 [astro-ph.HE].



\bibitem{SpadaPlanck}
  F.~Spada  [AMS-02 Collaboration],
  talk at the Tenth European Meeting From the Planck Scale to the Electroweak Scale (Planck 07), Warsaw, Poland, 9-13 Jun 2007.


\bibitem{Kounine}
  A.~Kounine,
  arXiv:1009.5349 [astro-ph.HE].



\bibitem{maestro}
  P.~Maestro,
  ``Indirect Search for Dark Matter by Measurements of the Cosmic Ray Positron Spectrum with the AMS-02 experiment,''
  PhD thesis, 2003.

\bibitem{bessp}
  S.~Haino {\it et al.},
  Phys.\ Lett.\  B {\bf 594} (2004) 35
  [arXiv:astro-ph/0403704].


\bibitem{Schmanau}
  M.~Schmanau,
{\it Prepared for the 29th International Cosmic Ray Conference (ICRC 2005), Pune, India, 3-11 Aug 2005}.

\bibitem{Casaus2009}
  J.~Casaus,
  J.\ Phys.\ Conf.\ Ser.\  {\bf 171} (2009) 012045.

\bibitem{fermianis}
  M.~Ackermann {\it et al.}  [Fermi LAT Collaboration],
  arXiv:1008.5119 [astro-ph.HE].

\bibitem{Ravi}
  V.~Ravi, R.~N.~Manchester, G.~Hobbs,
  Astrophys.\ J.\  {\bf 716} (2010) L85.
  [arXiv:1005.1966 [astro-ph.HE]].


\bibitem{Rankin}
  J.~M.~Ravi,
  Astrophys.\ J.\  {\bf 405} (1993) 285.
 
\bibitem{Watters}
  K.~P.~Watters, R.~W.~Romani, P.~Weltevrede and S.~Johnston,
  Astrophys.\ J.\  {\bf 695} (2009) 1289
  [arXiv:0812.3931 [astro-ph]].



\bibitem{Grasso2010}
  G.~Di Bernardo, C.~Evoli, D.~Gaggero, D.~Grasso, L.~Maccione and M.~N.~Mazziotta,
  arXiv:1010.0174 [astro-ph.HE].


\bibitem{cholis}
  D.~Malyshev, I.~Cholis and J.~Gelfand,
  Phys.\ Rev.\  D {\bf 80} (2009) 063005
  [arXiv:0903.1310 [astro-ph.HE]].




\end{thebibliography}
\end{document}